\documentclass[twocolumn]{aastex631}

\usepackage{amsmath}

\begin{document}

\title{Variable X-ray reverberation in the rapidly accreting AGN Ark 564: the response of the soft excess to the changing geometry of the inner accretion flow}

\author[0000-0003-0644-9282]{Zhefu Yu}
\affiliation{Kavli Institute for Particle Astrophysics and Cosmology, Stanford University, 452 Lomita Mall, Stanford, CA 94305, USA}

\author[0000-0002-4794-5998]{Dan Wilkins}
\affiliation{Kavli Institute for Particle Astrophysics and Cosmology, Stanford University, 452 Lomita Mall, Stanford, CA 94305, USA}

\author[0000-0003-0667-5941]{Steven W. Allen}
\affiliation{Kavli Institute for Particle Astrophysics and Cosmology, Stanford University, 452 Lomita Mall, Stanford, CA 94305, USA}
\affiliation{Department of Physics, Stanford University, 382 Via Pueblo Mall, Stanford, CA 94305, USA}
\affiliation{SLAC National Accelerator Laboratory, 2575 Sand Hill Road, Menlo Park, CA 94025, USA}



\begin{abstract}
\noindent X-ray reverberation, which exploits the time delays between variability in different energy bands as a function of Fourier frequency, probes the structure of the inner accretion disks and X-ray coronae of active galactic nuclei. We present a systematic X-ray spectroscopic and reverberation study of the high-Eddington-ratio narrow-line Seyfert 1 galaxy Ark 564, using over 900 ks of \textit{XMM-Newton} and \textit{NuSTAR} observations spanning 13 years. The time-averaged spectra can be well described by the a model consisting of a coronal continuum, relativistic disk reflection, warm Comptonization, and three warm absorbers. Leveraging the high X-ray brightness of Ark 564, we are able to resolve the time evolution of the spectra and contemporaneous reverberation lags. The soft-band lag relative to the continuum increases with the X-ray flux, while Fe K$\alpha$ lags are detected in only a subset of epochs and do not correlate with soft lags. Models based on a lamppost corona and reflection from a standard thin disk can broadly reproduce the observed lag-energy spectra of low-flux epochs; however, additional reverberation from the warm Comptonized atmosphere is required to explain the soft lags observed in high-flux epochs. A vertically puffed-up inner disk and a variable, vertically extended corona can better explain the observed evolution of the lags and covariance spectra. Our study underscores the importance of multi-epoch, multi-band analyses for a comprehensive understanding the inner accretion disk and corona.   
\end{abstract}



\section{Introduction} \label{sec:intro}
\noindent Active galactic nuclei (AGN), powered by luminous accretion flows onto supermassive black holes (SMBHs), commonly exhibit strong X-ray emission. The X-ray emission from AGN comprises at least two components produced by distinct physical mechanisms. First, a hot corona, with a temperature of $kT_e \sim 10^2$ keV, inverse Comptonizes ultraviolet (UV) photons from the accretion disk, generating a power-law continuum \cite[e.g.,][]{Eardley1975,Sunyaev1979}. Second, a fraction of the continuum photons irradiate the inner accretion disk and are reprocessed, producing a characteristic ``reflection'' spectrum. This reprocessing involves a combination of processes, including Compton scattering, fluorescence, and Bremsstrahlung emission from radiatively heated gas \citep[e.g.,][]{Guilbert1988,George1991,Ross2005}. Additionally, Doppler shifts and gravitational redshifts in the vicinity of the black hole can smear the reflection spectrum emerging from the inner disk. The characteristic features of a relativistic reflection spectrum include a soft X-ray excess, a broad Fe K$\alpha$ line near $\sim 6.4$ keV, and a Compton hump in the $\sim$ 20 - 40 keV range. 

In the simplest model, the corona is assumed to be a compact ``lamppost'' located on the spin axis of the SMBH \citep[e.g.,][]{Matt1991,Nayakshin2000}, and the accretion disk is modeled by a geometrically thin, optically thick disk \citep{Shakura1973}. However, this simplified picture may break down under certain conditions. For instance, at high accretion rates, radiation pressure generated within the disk causes the disk to be puffed up and form a ``slim'' disk \citep[e.g.,][]{Abramowicz1988,McKinney2012}. Moreover, lamppost geometry is adopted for mathematical simplicity, and the corona is likely to be spatially extended in reality \citep[e.g.,][]{Haardt1993}. 

One approach to measuring the properties of the corona and accretion disk is to fit relativistic reflection models to X-ray spectra. Early models of disk reflection, such as the \textsc{diskline} and \textsc{laor} models, were limited to modeling only the Fe K$\alpha$ line \citep{Fabian1989,Laor1991}. More recent models, such as \textsc{reflionx} \citep{Ross2005} and \textsc{xillver} \citep{Garcia2010,Garcia2013}, predict the full non-relativistic reflection spectrum with an improved treatment of atomic processes. The \textsc{relxill} model self-consistently incorporates general relativistic blurring to the reflection spectrum predicted by \textsc{xillver}, accounting for the angular dependence of the reflection \citep{Garcia2014,Dauser2014}. Fitting reflection models to X-ray spectra enables constraints on key parameters, including the black hole spin, disk inclination, ionization parameter, density, and the properties of the corona.

However, the robustness of X-ray spectral constraints can be compromised by degeneracies between different physical mechanisms that can produce similar spectral features. For example, Comptonization of UV photons by a warm plasma at a temperature of $\sim 0.1$ - $1$ keV spanning the inner accretion flow can also generate excess soft X-ray emission \citep[e.g.,][]{Czerny2003,Done2012,Ehler2018,Tripathi2019}. Indeed, previous studies have found that the X-ray spectra of some AGN cannot be fully described by disk reflection alone, requiring the addition of such a warm Comptonized component \citep[e.g.,][]{Porquet2018,Taylor2025}.

Another method for probing the corona and the innermost regions of accretion flows is X-ray reverberation \citep[for recent reviews see][]{Uttley2014,Cackett2021}. AGN X-ray emission typically exhibits strong temporal variability, and X-ray reverberation exploits the time delay between variations in the coronal continuum and disk reflection spectrum. Moreover, AGN variability can be dominated by different physical processes on different time scales. To disentangle these time scales, X-ray time lags are commonly measured in the Fourier domain as a function of frequency. At relatively high frequencies, corresponding to the variability at short time scales, the disk reflection spectrum lags behind the coronal continuum. This reverberation lag is governed by the light travel time in the curved spacetime from the corona to the disk, thereby encoding critical information on the black hole spin and the structure of both the corona and the disk.

The first detection of X-ray reverberation lag was reported for 1H0707-495, where the soft band dominated by the soft excess lagged behind the hard continuum band at a frequency of $\sim 10^{-3}$ Hz, supporting the disk reflection scenario and indicating that at least part of the soft excess in this system originates from reflection \citep{Fabian2009}. Subsequent studies have measured reverberation lags associated with the Fe K$\alpha$ line relative to the continuum-dominated bands \citep[e.g.,][]{Zoghbi2012,Kara2016}. In contrast, at lower frequencies the hard band has been observed to lag behind the soft band in both AGN and galactic black hole binaries, although the physical origin of this low-frequency lag remains uncertain \citep[e.g.,][]{Miyamoto1989,Vaughan2003,Kara2013,Kara2016}. Given the substantial photon counts required, most X-ray reverberation studies have relied on stacking multiple observations to derive an averaged time lag, thus limiting the ability to trace the temporal evolution of the disk and corona. 

In this paper, we focus on the narrow-line Seyfert 1 (NLS1) galaxy Ark 564 at a redshift of $z = 0.0247$. Ark 564 is among the brightest X-ray AGN and is accreting near the Eddington limit \citep[e.g.,][]{Zhang2006,Kara2017}. Extensive X-ray spectroscopic studies have been performed on this source \citep[e.g.,][]{Giustini2015,Khanna2016,Jiang2019,Lewin2022,Lyu2024}. The time averaged X-ray spectrum of Ark 564 exhibits a prominent soft excess typical of NLS1 galaxies. \citet{Giustini2015} and \citet{Khanna2016} analyzed \textit{XMM-Newton} Reflection Grating Spectrometer (RGS) spectra, identifying three distinct warm absorbers with different ionization states. \citet{Kara2017} presented the first \textit{NuSTAR} observation of Ark 564, showing that it harbors one of the coolest X-ray corona among AGN. \citet{McHardy2007}, \citet{Legg2012}, and \citet{Kara2013} reported measurements of high-frequency soft lags using \textit{XMM-Newton} observations from 2005 and 2011, with \citet{Kara2013} additionally detecting high-frequency Fe K$\alpha$ lags and low-frequency hard lags. \citet{Lewin2022} employed a Gaussian process approach to jointly model the spectra and lags from simultaneous \textit{XMM-Newton} and \textit{NuSTAR} observations made in 2018. All of these studies have either focused on a single epoch or provided averaged lag measurements over multiple epochs, without tracing the temporal evolution.

We present a systematic X-ray spectral and reverberation analysis of Ark 564. We jointly model the X-ray spectra from over 900 ks of \textit{XMM-Newton} \citep{xmmnewton} and \textit{NuSTAR} \citep{nustar} observations obtained over 13 years. We then resolve the X-ray lags for individual observations and investigate their temporal evolution. The paper is organized as follows. In Section \ref{sec:obs}, we describe the observations and data reduction. Section \ref{sec:spec} details the spectroscopic analysis. We present the lag measurements in Section \ref{sec:rm} and the lag models based on ray-tracing simulations in Section \ref{sec:sim}. Section \ref{sec:discussion} further discusses the evolution of the lags in the context of a slim disk and extended corona. Finally, we summarize our findings in Section \ref{sec:summary}.

\section{Observations} \label{sec:obs}

\begin{figure*}
\gridline{\fig{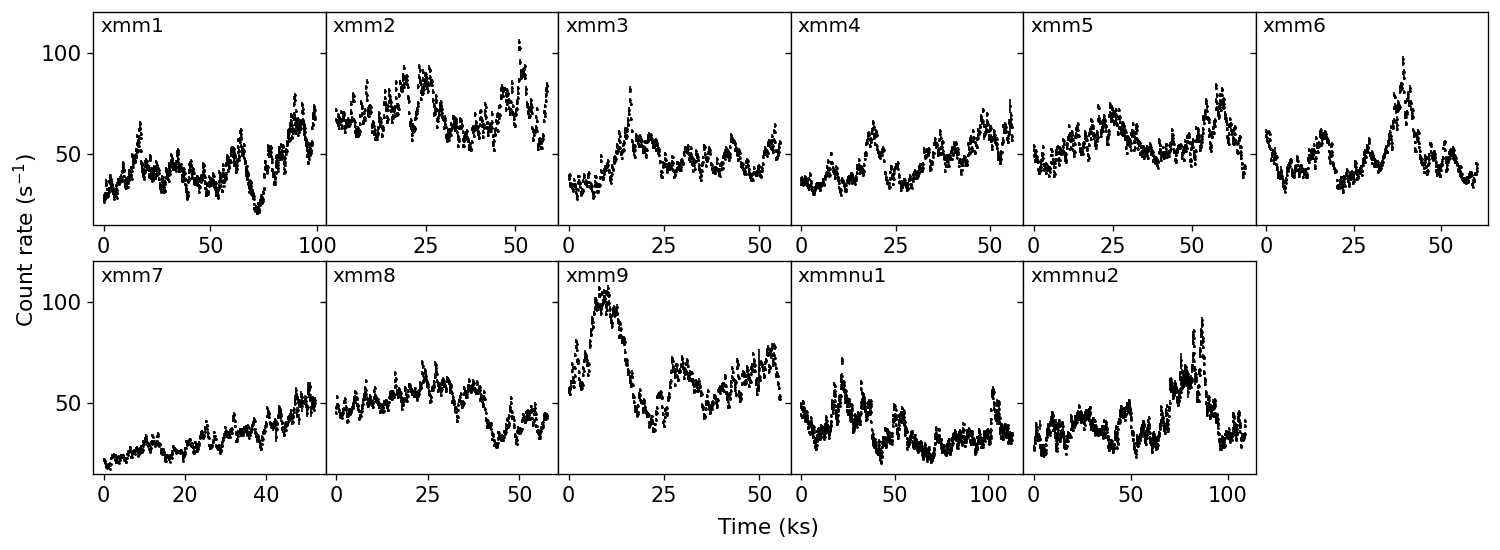}{\textwidth}{}}
\figcaption{X-ray light curves from \textit{XMM-Newton} observations in the 0.3 - 10 keV band. The light curves are binned in 100 s for visibility, while finer binning is employed in the timing analysis. The epoch name is given in the upper left corner of each panel. \label{fig:lc}}
\end{figure*}

\noindent We use the archival \textit{XMM-Newton} European Photon Imaging Camera (EPIC)-pn and \textit{NuSTAR} observations listed in Table \ref{tab:obs}. \textit{XMM-Newton} observations were reduced using the \textsc{xmm science analysis system (sas)} v21.0.0. The event lists were reprocessed using the \textsc{epproc} task, applying the latest available version of the calibration. The source and background spectra were extracted using the \textsc{evselect} task and the corresponding response matrices and ancillary response (\textit{i.e.} effective area) files were generated using \textsc{rmfgen} and \textsc{arfgen}. Light curves were extracted from the \textit{XMM-Newton} observations, also using \textsc{evselect}, and were corrected to account for dead time and exposure variations using the \textsc{epiclccorr} task. For the extraction of both spectra and light curves, we define a circular source extraction region, 35\,arcsec in diameter, centered on the point source, and a corresponding background extraction region same size, located on the same detector chip. Figure \ref{fig:lc} shows the light curves from the \textit{XMM-Newton} observations in 0.3 - 10 keV. The baseline of the light curves is about 100 ks for the epochs xmm1, xmmnu1, xmmnu2 and about 50 ks for the other epochs. 

\citet{Jethwa2015} provides metrics to characterize the effect of pile-up on \textit{XMM-Newton} observations. For EPIC-pn observations in small window mode, a count rate of 50 s$^{-1}$ is considered the ``tolerant limit'', corresponding to a 1-1.5\% distortion in the spectral shape due to pile-up. All the \textit{XMM-Newton} observations in our analysis fall within this limit except for two epochs, xmm2 and xmm9. We estimate a spectral distortion of $\sim 0.7\%$ and $\sim 1.4\%$ for the lowest flux epoch xmm7 and the highest flux epoch xmm2, respectively, based on Figure 5 of \citet{Jethwa2015} and the spectral shape of Ark 564. We enable the pile-up correction option when generating the rmf files using \textsc{rmfgen}, which simulates charges by randomizing over the detector response and calculates a corrected response matrix by comparing the event reconstruction results with and without the simulated charges added to the observed data \footnote{https://xmm-tools.cosmos.esa.int/external/sas/current/ \newline doc/rmfgen/node14.html}. Although this simulation does not fully reproduce the charge distribution caused by pile-up, its impact on our analysis should be minimal given the insignificant level of pile-up. Overall, pile-up should not have a significant effect on our results. 

\textit{NuSTAR} observations were reduced using the \textit{NuSTAR} data-analysis system, \textsc{nustardas}, v2.1.4. The event lists from each of the focal plane module (FPM) detectors were reprocessed and filtered using the \textsc{nupipeline} task, applying the most recent calibration available at the time of writing. Source and background spectra were extracted using the \textsc{nuproducts} task, defining a circular source extraction region, 60\,arcsec in diameter, centered on the point source, and a corresponding background extraction region same size, located on the same detector chip.

Additional X-ray observations of Ark 564 exist but are not included in this work. \textit{XMM-Newton} conducted four observations in 2000 and 2001 (observation IDs from 0006810101 to 0006810401). The $\sim 7$ ks exposure times of 0006810101 and 0006810301 are insufficient for reliable reverberation analysis, while 0006810201 and 0006810401 do not have EPIC observations. Two additional \textit{NuSTAR} epochs were obtained in May 2015 and June 2018 (observation IDs 60101031002, 60401031002), but their lack of temporal overlap with the \textit{XMM-Newton} data limits their utility in providing spectral constraints. We also exclude a 2015 Suzaku observation (observation ID 710018010) due to the complexity of its background and the impact of pile-up.

\begin{deluxetable}{lllcc}
\tablecaption{The \textit{XMM-Newton} and \textit{NuSTAR} observations analyzed in this work. Column (1) lists a short name for each observation, where the names beginning with ``xmm'' and ``nu'' correspond to \textit{XMM-Newton} and \textit{NuSTAR} data, respectively. Columns (2) - (3) give the observation ID and the start date of each epoch. Column (4) provides the clean exposure time. For \textit{NuSTAR}, the exposure time is the average of FPMA and FPMB. Column (5) gives the count rate, measured in the 0.3 - 10 keV band for \textit{XMM-Newton} and in the 3 - 27 keV band for \textit{NuSTAR}, without background subtraction. For \textit{NuSTAR}, the count rate represents the sum of FPMA and FPMB. \label{tab:obs}}
\tabletypesize{\footnotesize}
\tablehead{
\colhead{Name} &\colhead{ObsID} & \colhead{Start Date} & \colhead{Exposure} & \colhead{Count rate} \\
  &  &  & \colhead{(ks)} & \colhead{(s$^{-1}$)} 
}
\startdata
xmm1 & 0206400101 & 2005-01-05 & 69 & 34.4 \\
xmm2 & 0670130201 & 2011-05-24 & 41 & 55.2 \\
xmm3 & 0670130301 & 2011-05-30 & 39 & 36.6 \\
xmm4 & 0670130401 & 2011-06-05 & 38 & 37.2 \\
xmm5 & 0670130501 & 2011-06-11 & 47 & 44.3 \\
xmm6 & 0670130601 & 2011-06-17 & 41 & 38.6 \\
xmm7 & 0670130701 & 2011-06-25 & 33 & 25.0 \\
xmm8 & 0670130801 & 2011-06-29 & 40 & 38.5 \\
xmm9 & 0670130901 & 2011-07-01 & 39 & 50.8 \\
xmmnu1 & 0830540101 & 2018-12-01 & 78 & 29.1 \\
xmmnu2 & 0830540201 & 2018-12-03 & 67 & 32.7 \\
\hline
nu1 & 60401031004 &  2018-11-28 & 408 & 0.61 \\
\enddata
\end{deluxetable}

\section{Spectroscopic Analysis} \label{sec:spec}
\noindent We bin the \textit{XMM-Newton} and \textit{NuSTAR} spectra using the \textsc{grppha} task so that there is at least one count per energy bin. We use \textsc{XSPEC} version 12.14.1 \citep{xspec} to analyze the spectra, employing the modified $C$-statistic \citep{Cash1979} for the case when the background spectrum is subtracted, also known as the $W$-statistic. 

\subsection{Simultaneous NuSTAR and XMM-Newton Epochs} \label{subsec:spec_xmnu}
\noindent We began our analysis by constructing a model to fit the broadband spectra of three epochs xmmnu1, xmmnu2, and nu1, obtained by joint \textit{XMM-Newton} and \textit{NuSTAR} observations in 2018. The two neighboring \textit{XMM-Newton} epochs xmmnu1 and xmmnu2 were obtained between December 1 and 3, coinciding with the \textit{NuSTAR} epoch nu1 obtained from November 28 to December 8. We modeled the accretion disk reflection using the relativistic reflection model \textsc{relxilllpCp} \citep{Garcia2014,Dauser2014}, which differs from other \textsc{relxill}-type models in two principal aspects: First, it models the coronal continuum with the physically motivated Comptonization model \textsc{nthcomp} \citep{nthcomp_Zdziarski1996,nthcomp_Zycki1999} rather than an empirical power law with an exponential cutoff. Second, it self-consistently computes the disk emissivity profile, defined as the flux reflected by the disk as a function of radius, under the assumption of a lamppost geometry (while the default \textsc{relxill} model adopts an empirical broken power law for the emissivity). The free parameters of the \textsc{relxilllpCp} model include the black hole spin $a$\footnote{The black hole spin parameter is defined as $a=J/Mc$ in unit of $GM/c^2$, where $J$ and $M$ are the angular momentum and mass of the black hole, respectively.}, the corona height $h$, the disk inclination, the disk density log($n_e$), the iron abundance $A_{\rm Fe}$, the ionization parameter log($\xi$), and the normalization. 

We disabled the coronal continuum calculation in \textsc{relxilllpCp} model by fixing the reflection fraction to $-1$ and included a separate continuum component. To remain consistent with \textsc{relxilllpCp}, we employed \textsc{nthcomp} to model the coronal continuum. The free parameters of \textsc{nthcomp} are the power-law slope $\Gamma$, the electron temperature $kT_e$, and the normalization. We tied the values of $\Gamma$ and $kT_e$ to those in the \textsc{relxilllpCp} model for self-consistency. This \textsc{relxilllpCp} + \textsc{nthcomp} model is equivalent to a single \textsc{relxilllpCp} model with a free reflection fraction, but the two-component approach facilitates the separation of the coronal continuum and disk reflection in \textsc{XSPEC}. 

We found that the reflection from the disk alone was unable to describe the strong soft excess in Ark 564, and we need to include an additional soft excess component \textsc{comptt} described by the Comptonization of UV photons by a warm ($\sim 0.15$ keV) plasma. This \textsc{comptt} component has been commonly used to model the soft excess, which is likely associated with a warm atmosphere on the accretion disk \citep[e.g.,][]{Czerny2003,Middleton2009,Done2012}.

Previous studies of the X-ray absorption in Ark 564 identified three distinct warm absorber components based on the RGS spectra of \textit{XMM-Newton} epochs xmm1 - xmm9 \citep{Giustini2015,Khanna2016}. Motivated by these results, we model the absorption using a combination of the interstellar medium absorption model \textsc{tbabs} and three warm absorber components. We fixed the \textsc{tbabs} column density to the Galactic value of $5.3 \times 10^{20} \, {\rm cm}^{-2}$ \citep{Galactic_nH}. We generated table models for the warm absorbers using \textsc{XSTAR} version 2.59, with configurations similar to those in \citet{Giustini2015}. In our models, we adopted solar abundances, a turbulence velocity of 100 km s$^{-1}$, and a power-law ionizing continuum with an index of $\Gamma=2.5$. We constructed a grid for the ionization parameter log($\xi$) with 25 linearly sampled points between $-$1.5 and 3.5, and a grid for the column density $N_{\rm H}$ with 15 logarithmically sampled points spanning $5 \times 10^{18} \, {\rm cm}^{-2}$ to $10^{21} \, {\rm cm}^{-2}$. In addition to the multiplicative absorption tables, we included the transmitted emission models from \textsc{XSTAR} as additive components. 
 
\begin{figure}
\epsscale{1.15}
\plotone{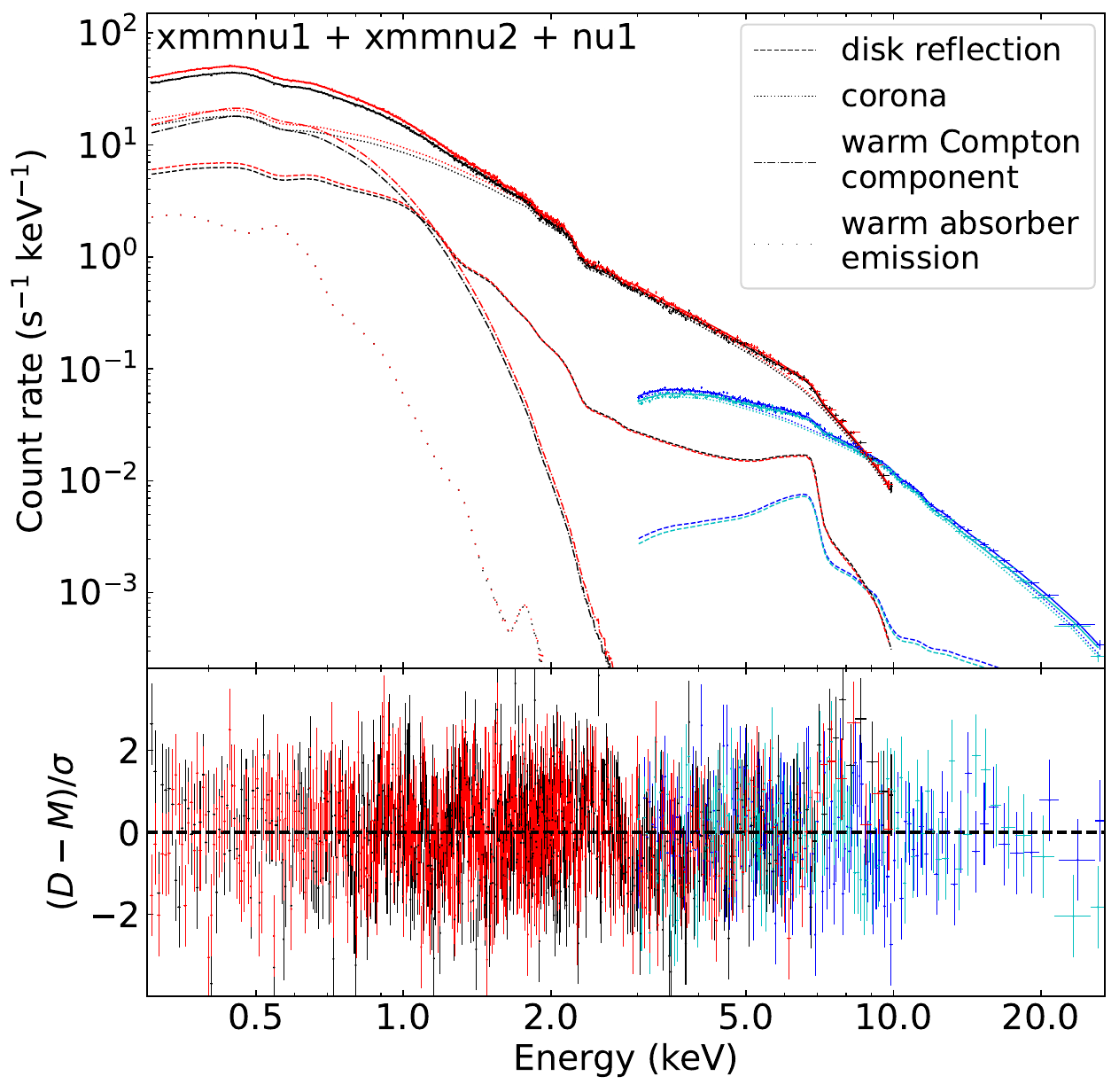}
\figcaption{\textit{(upper panel)} X-ray spectra and best-fit models for xmmnu1 (black), xmmnu2 (red), nu1 FPMA (blue), and nu1 FPMB (cyan). The solid crosses denote the observed spectra. The dashed, dotted, dash-dotted and sparsely dotted lines represent the disk reflection \textsc{relxilllpCp}, the coronal continuum \textsc{nthcomp}, the warm Compton component \textsc{comptt}, and the transmitted emission from the warm absorbers modeled by the XSTAR additive tables, respectively. Only one XSTAR additive component is sufficiently significant to be visible in this plot. \textit{(lower panel)} Residuals of the fit, defined as the difference between the data and the model divided by the uncertainty. \label{fig:specfit}}
\end{figure}

The full spectral model is given by
\begin{multline}
{\rm model} = {\textsc{constant}} \times {\textsc{tbabs}} \times {\textsc{xs1}} \times {\textsc{xs2}} \times {\textsc{xs3}} \times ({\textsc{xs1a}} \\
  + {\textsc{xs2a}} + {\textsc{xs3a}} + {\textsc{relxilllpCp}} + {\textsc{nthcomp}} + {\textsc{comptt}})
\label{eq:model}
\end{multline}
where the constant is for the cross-calibration between the flux measured by different instruments, the multiplicative components \textsc{xs1} - \textsc{xs3} represent the warm absorbers, and the additive components \textsc{xs1a} - \textsc{xs3a} denote the transmitted emission from each warm absorber. We jointly fit the FPMA and FPMB spectra of the \textit{NuSTAR} epoch nu1 together with the \textit{XMM-Newton} epochs xmmnu1 and xmmnu2 as four data groups in \textsc{XSPEC}. The constant is fixed to unity for all groups except for the nu1 FPMB spectra; for FPMB, we allow the constant to vary to account for the calibration differences between FPMA and FPMB, while all other parameters of FPMB are tied to FPMA. For xmmnu1 and xmmnu2, we allow the power-law slope $\Gamma$ and the normalizations of the \textsc{relxilllpCp}, \textsc{nthcomp}, and \textsc{comptt} components to vary, while the remaining parameters are tied to those of nu1. Tables \ref{tab:specpar} and \ref{tab:specpar_xs} summarize the parameter links among different epochs. This model yields a reasonable fit with a modified $C$-statistic value of $C = 5475$ for a degree of freedom of ${\rm dof} = 5050$. Figure \ref{fig:specfit} shows the spectral fit and residuals. The residuals do not exhibit significant features except for a small bump in the 7 - 9 keV range in the \textit{XMM-Newton} data. \citet{Lewin2022} observed similar residuals with different models. Since the \textit{XMM-Newton} instrumental background in this energy range exhibits several bright lines, and since the \textit{NuSTAR} data do not display significant residuals in the same energy range, we do not consider this feature further. 

We also explored several modifications to this model. Excluding all warm absorbers leads to a significantly poorer fit with $C / {\rm dof} = 5785 / 5059$. We did not include a fourth warm absorber, as it produced only a marginal improvement with $C / {\rm dof} = 5471 / 5047$. For the disk reflection component, we tried replacing the \textsc{relxilllpCp} model with \textsc{relxillCp}, which employs an empirical broken power law for the disk emissivity profile; this alternative yielded similar fit statistics $C / {\rm dof} = 5473 / 5048$. Substituting the non-relativistic reflection model \textsc{xillverCp} results in a substantially worse fit with $C / {\rm dof} = 5642 / 5054$. Removing the warm Compton component \textsc{comptt} resulted in a significantly worse fit with $C / {\rm dof} = 7649 / 5054$, indicating that the disk reflection alone cannot fit the data well even when the disk density is allowed to vary. Finally, we tested the \textsc{reXcor} model which simultaneously calculates the disk reflection and warm Comptonization \citep[e.g.,][]{Xiang2022,Ballantyne2024}. For assumed coronal heights of $5r_g$ and $20r_g$, we obtained significantly poorer fits with $C / {\rm dof} = 5934 / 5056$ and $5930 / 5056$, respectively. Overall, we did not identify a combination of models that substantially improves upon Equation (\ref{eq:model}), so we adopt it as the fiducial spectral model for the remainder of the paper.

\subsection{Joint Fitting of All Epochs} \label{subsec:spec_all}
\noindent We then performed a joint fit to the data from all epochs using the fiducial model. We tied the disk inclination, black hole spin parameter $a$, disk density log($n_e$), and iron abundance $A_{\rm Fe}$ across different epochs, while allowing the other parameters to vary. We obtain a reasonable fit with $C / {\rm dof} = 23709 / 22304$. Tables \ref{tab:specpar} and \ref{tab:specpar_xs} list the best-fit model parameters and their 1$\sigma$ (68.3\%) statistical uncertainties. Spectral fit plots for individual epochs are provided in Appendix \ref{appsec:figs}. For epochs xmmnu1, xmmnu2, and nu1, the model parameters are consistent with those derived in Section \ref{subsec:spec_xmnu} at $1\sigma$ level, except for the inclination which exhibits a slightly larger difference of $\sim 1.5\sigma$.  

For xmmnu1 and xmmnu2, we obtain a corona temperature of $kTe \sim 55$ keV, which is higher than that reported by \citet{Kara2017}. To investigate this discrepancy, we fitted the 2015 NuSTAR epoch used by \citet{Kara2017} (observation ID 60101031002) with \textsc{relxilllpCp} and obtained $kTe \sim 47$ keV, which is still higher than their reported value. One likely explanation is that \citet{Kara2017} employed the non‐relativistic reflection model \textsc{xillverCp}, which tends to yield a lower coronal temperature than \textsc{relxilllpCp} for the same spectral shape. The temperature $kT_e$ is poorly constrained for xmm1 - xmm9 due to the lack of hard X-ray data. We obtain large corona heights on the order of several tens of $r_g$ with significant uncertainties in most epochs, except for xmmnu1 and xmmnu2 where the corona height is smaller and tightly constrained. For xmm1 and xmm4, we report only the $2\sigma$ limits on corona height. The ionization parameter ${\rm log}(\xi)$ is high for all epochs, with xmm9 exhibiting ${\rm log}(\xi)$ pegged at the maximum value. 

The joint fitting yields improved constraints on the disk inclination and the black hole spin parameter $a$. The disk inclination of $38.4^{+1.2}_{-1.1}$ degrees is similar to the results of \citet{Lewin2022} and falls within the broad range of values reported by \citet{Kara2017} using different models. We find a high but non-maximal spin $a=0.70^{+0.07}_{-0.11}$, in contrast to previous studies that either pegged the spin at its maximum value or obtained poorly constrained estimates \citep{Walton2013,Kara2017,Tripathi2018,Jiang2019}. The disk density is pegged at its lower boundary of ${\rm log}N = 15$, similar to \citet{Lewin2022} where the 1$\sigma$ lower limit of log($n_e$) reaches this boundary. One possible explanation is that the inner disk becomes dominated by radiation pressure, which alters its radiative transfer properties and yields an effective density that differs from that of the outer disk. Although we can obtain larger values of log($n_e$) as reported in previous studies \citep{Jiang2019,Lyu2024} by removing the \textsc{comptt} component, this approach results in a significantly poorer fit (see Section \ref{subsec:spec_xmnu}). The iron abundance $A_{\rm Fe}$ is pegged at its maximum value, reflecting the limitations of our data and model in constraining this parameter. Similar high iron abundances have been reported in other sources, such as 1H 1934$-$063 and IRAS 13224$-$3809 \citep[e.g,][]{Frederick2018,Jiang2018}. 

The warm absorbers exhibit a combination of low (${\rm log}(\xi) < 0$) and intermediate (${\rm log}(\xi) \sim$ 0.8 - 1.6) ionization states in most epochs, while a few epochs also require a high-ionization warm absorber (${\rm log}(\xi) > 2$). In some epochs, \textsc{XSPEC} employs two XSTAR models to represent the same warm absorber and the best fit is obtained when the ionization parameters of two components are equal, indicating that only one or two distinct absorbers are needed. This range of ionization states is qualitatively consistent with the results of \citet{Giustini2015}. The transmitted emission has negligible contribution to the fit in most cases. The weak constraints on the warm absorber parameters are expected given the low energy resolution of the \textit{XMM-Newton} EPIC-pn instrument. Therefore, we interpret these parameters as only a phenomenological prescription to be marginalized over, rather than as precise physical constraints on the warm absorbers.

\begin{figure}
\epsscale{1.2}
\plotone{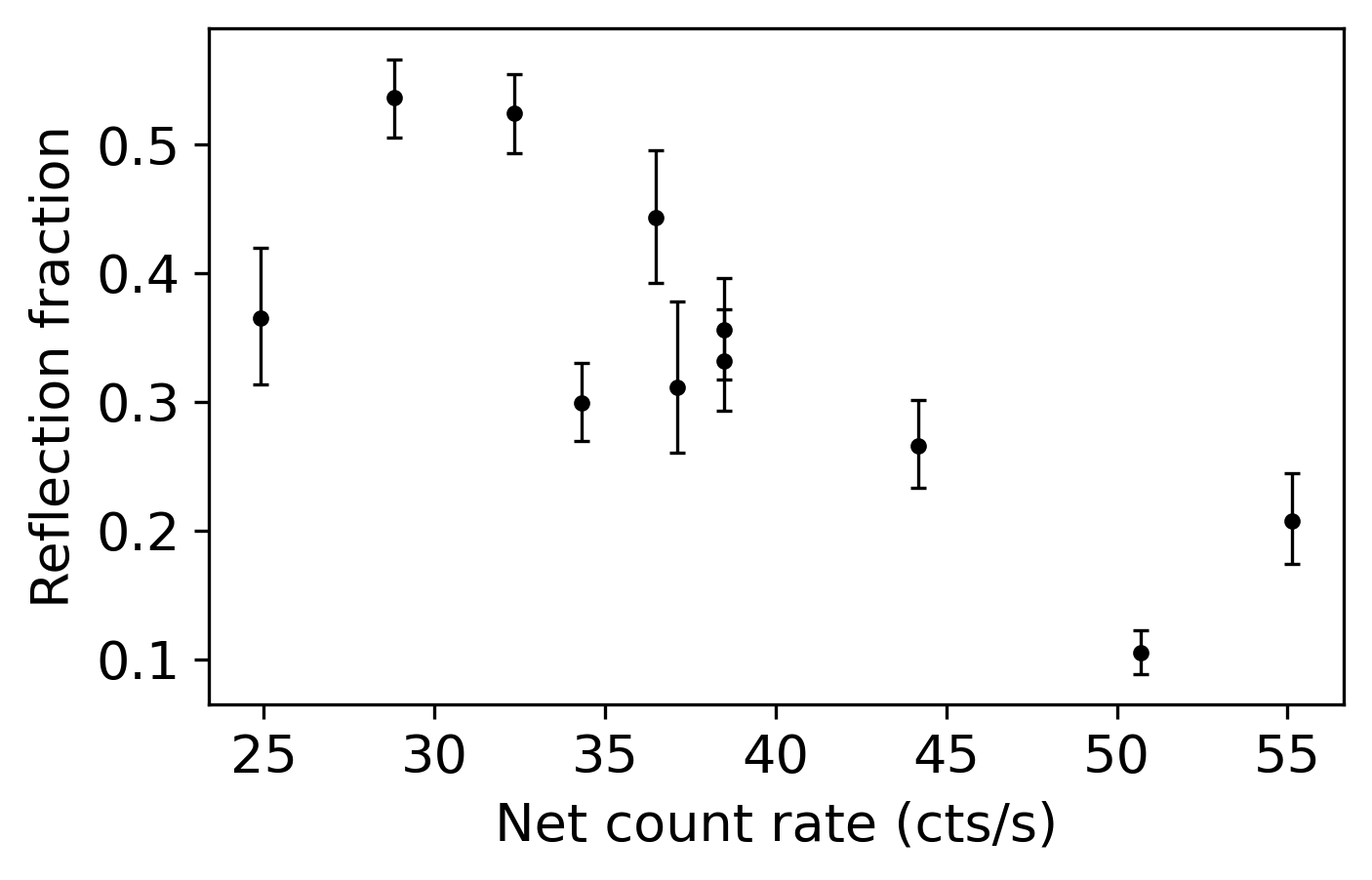}
\figcaption{Reflection fraction, defined as the ratio of the coronal continuum reflected by the disk to that emitted directly towards the observer, as a function of net count rate. \label{fig:rffr}}
\end{figure}

Our fiducial model does not directly provide the reflection fraction, defined as the ratio of the coronal continuum reflected by the disk to that emitted directly towards the observer, since it is fixed at $-1$ to disable the coronal continuum in the \textsc{relxilllpCp} model. To constrain this parameter, we removed the \textsc{nthcomp} component and allowed the reflection fraction in \textsc{relxilllpCp} to vary freely. Figure \ref{fig:rffr} shows the reflection fraction as a function of net count rate. The reflection fraction is low in all epochs and exhibits a decreasing trend with increasing count rate. Similar anti-correlations have been reported in other sources, such as MCG–6-30-15 and Mrk 335 \citep[e.g.,][]{Miniutti2004,Wilkins2015_reffrac}.

\begin{deluxetable*}{c|rrr|rrr|rrrrcccc}
\tablecaption{Model parameters from the joint fit of all \textit{XMM-Newton} and \textit{NuSTAR} epochs. Column (1) gives the short name of each epoch. Columns (2) - (4) give the power-law slope $\Gamma$, the electron temperature $kT_e$ (unit: keV), and the normalization of the \textsc{nthcomp} model. Columns (5) - (7) give the electron temperature $kT_e$ (unit: keV), the optical depth $\tau_p$, and the normalization of the \textsc{comptt} model. Columns (8) - (14) give the corona height $h$ (unit: gravitational radius $r_g$), the disk ionization parameter log($\xi$), the normalization, the disk inclination (unit: degree), the black hole spin parameter $a$, the disk density log($n_e / {\rm cm}^{-3}$), and the iron abundance $A_{\rm Fe}$ (unit: solar abundance) of the \textsc{relxilllpCp} model. The model normalization is expressed as the photon flux $F_x$ (unit: $10^{-3}$ cm$^{-2}$ s$^{-1}$) in the 0.3 - 10 keV band for the \textit{XMM-Newton} epochs and in the 3 - 10 keV band for the \textit{NuSTAR} epoch. Since the \textsc{comptt} model contributes negligibly above 3 keV, we fix its photon flux to 0 for the \textit{NuSTAR} epoch nu1. The corona height $h$ for xmm1 is high with a large uncertainty, so we report only its $2\sigma$ lower limit. We also report the $2\sigma$ bounds for parameters where we can only constrain an upper or lower limit. \label{tab:specpar}}
\tabletypesize{\scriptsize}
\tablehead{\multicolumn{1}{c|}{} & \multicolumn{3}{c|}{\textsc{nthcomp}} & \multicolumn{3}{c|}{\textsc{comptt}} & \multicolumn{7}{c}{\textsc{relxilllpCp}}  \\
\cline{2-14}
\multicolumn{1}{c|}{Name} & \colhead{$\Gamma$}  & \colhead{$kT_e$}  & \multicolumn{1}{c|}{$F_x$} & \colhead{$kT_e$} & \colhead{$\tau_p$} & \multicolumn{1}{c|}{$F_x$} & \colhead{$h$} & \colhead{log($\xi$)} & \colhead{$F_x$} & \colhead{Incl.} & \colhead{$a$} & \colhead{log($n_e$)} & \colhead{$A_{\rm Fe}$} 
}
\startdata
xmm1 & $2.504^{+0.004}_{-0.007}$ & $> 26$ & $55.6^{+0.5}_{-0.6}$ & $0.161^{+0.005}_{-0.003}$ & $19.7^{+1.0}_{-1.7}$ & $57.1^{+2.7}_{-2.0}$ & $> 31.6$ & $3.03^{+0.03}_{-0.03}$ & $8.1^{+0.9}_{-0.9}$ & - & - & - & - \\
xmm2 & $2.591^{+0.006}_{-0.006}$ & $> 30$ & $98.6^{+1.0}_{-1.0}$ & $0.163^{+0.004}_{-0.005}$ & $22.5^{+1.3}_{-1.2}$ & $56.6^{+2.6}_{-3.7}$ & $22.1^{+15.7}_{-10.1}$ & $3.27^{+0.05}_{-0.06}$ & $13.7^{+2.6}_{-2.2}$ & - & - & - & - \\
xmm3 & $2.558^{+0.008}_{-0.008}$ & $> 26$ & $58.4^{+0.9}_{-0.7}$ & $0.148^{+0.004}_{-0.003}$ & $28.7^{+3.3}_{-2.0}$ & $57.4^{+3.0}_{-3.8}$ & $14.7^{+8.8}_{-5.5}$ & $2.98^{+0.05}_{-0.09}$ & $11.0^{+2.1}_{-2.6}$ & - & - & - & - \\
xmm4 & $2.479^{+0.006}_{-0.016}$ & $> 193$ & $54.8^{+0.6}_{-0.6}$ & $0.199^{+0.013}_{-0.012}$ & $14.7^{+2.0}_{-1.7}$ & $44.2^{+1.8}_{-2.6}$ & $< 13.3$ & $3.27^{+0.06}_{-0.10}$ & $10.3^{+1.9}_{-1.3}$ & - & - & - & - \\
xmm5 & $2.541^{+0.006}_{-0.007}$ & $> 23$ & $72.9^{+0.4}_{-0.4}$ & $0.156^{+0.005}_{-0.005}$ & $22.4^{+1.6}_{-1.6}$ & $56.2^{+0.6}_{-2.4}$ & $31.0^{+31.2}_{-14.6}$ & $3.22^{+0.06}_{-0.05}$ & $12.6^{+1.2}_{-1.1}$ & - & - & - & - \\
xmm6 & $2.519^{+0.006}_{-0.006}$ & $> 36$ & $62.5^{+0.7}_{-0.9}$ & $0.162^{+0.010}_{-0.003}$ & $20.3^{+1.5}_{-2.5}$ & $48.9^{+1.0}_{-1.9}$ & $20.3^{+9.2}_{-6.5}$ & $3.04^{+0.05}_{-0.04}$ & $9.8^{+0.5}_{-0.4}$ & - & - & - & - \\
xmm7 & $2.473^{+0.008}_{-0.012}$ & $> 14$ & $35.2^{+0.6}_{-0.5}$ & $0.164^{+0.004}_{-0.006}$ & $21.4^{+0.7}_{-1.6}$ & $34.1^{+3.6}_{-2.6}$ & $29.5^{+16.9}_{-11.8}$ & $3.13^{+0.06}_{-0.05}$ & $7.4^{+0.9}_{-0.9}$ & - & - & - & - \\
xmm8 & $2.524^{+0.005}_{-0.008}$ & $> 20$ & $60.9^{+0.6}_{-0.7}$ & $0.134^{+0.004}_{-0.003}$ & $34.1^{+4.7}_{-4.4}$ & $48.3^{+1.6}_{-1.9}$ & $22.6^{+10.8}_{-10.6}$ & $3.22^{+0.06}_{-0.05}$ & $13.8^{+1.3}_{-1.3}$ & - & - & - & - \\
xmm9 & $2.599^{+0.009}_{-0.005}$ & $> 37$ & $82.7^{+1.3}_{-1.7}$ & $0.156^{+0.005}_{-0.004}$ & $25.0^{+2.0}_{-1.5}$ & $65.2^{+4.3}_{-2.8}$ & $18.9^{+14.5}_{-7.8}$ & $> 4.51$ & $11.0^{+1.8}_{-1.9}$ & - & - & - & - \\
\hline
xmmnu1 & $2.496^{+0.007}_{-0.006}$ & - & $42.2^{+0.4}_{-0.5}$ & - & - & $30.1^{+1.9}_{-1.6}$ & - & - & $14.7^{+1.6}_{-1.2}$ & - & - & - & - \\
xmmnu2 & $2.524^{+0.008}_{-0.006}$ & - & $47.6^{+0.4}_{-0.3}$ & - & - & $35.6^{+1.2}_{-2.2}$ & - & - & $16.1^{+1.3}_{-1.4}$ & - & - & - & - \\
nu1 & $2.513^{+0.013}_{-0.011}$ & $55^{+81}_{-15}$ & $1.46^{+0.01}_{-0.01}$ & $0.141^{+0.005}_{-0.005}$ & $29.7^{+2.9}_{-2.3}$ & 0 & $5.5^{+1.2}_{-1.5}$ & $3.28^{+0.04}_{-0.03}$ & $0.14^{+0.01}_{-0.01}$ & $38.4^{+1.2}_{-1.1}$ & $0.70^{+0.07}_{-0.11}$ & $< 15.04$ & $> 9.6$
\enddata
\end{deluxetable*}

\begin{deluxetable*}{c|rrr|rrr|rrr}
\tablecaption{Warm absorber parameters from the joint fit of all \textit{XMM-Newton} and \textit{NuSTAR} epochs. Columns (2) - (3) list the ionization parameter log($\xi$) and the column density $N_H$ (unit: $10^{20} \, {\rm cm}^{-2}$) for the first warm absorber component, while Column (4) gives the photon flux $F_x$ (unit: $10^{-3}$ cm$^{-2}$ s$^{-1}$) of the transmitted emission from this component. The subsequent columns give the corresponding parameters for the second and third warm absorber components. We report only the $2\sigma$ bounds for parameters where we can only constrain an upper or lower limit. The ``unc.'' markers denote either the transmitted emission flux $F_x$ that is extremely small and has a negligible impact on the fit, or other parameters that we are unable to constrain within the allowed parameter space. The XSTAR parameters are treated only as phenomenological prescriptions to be marginalized over, rather than as precise physical constraints on the warm absorbers. \label{tab:specpar_xs}}
\tabletypesize{\scriptsize}
\tablehead{\multicolumn{1}{c|}{} & \multicolumn{3}{c|}{\textsc{XSTAR 1}} & \multicolumn{3}{c|}{\textsc{XSTAR 2}} & \multicolumn{3}{c}{XSTAR 3} \\
\cline{2-10}
\multicolumn{1}{c|}{Name} & \colhead{log($\xi$)} & \colhead{$N_H$} &  \multicolumn{1}{c|}{$F_x$} & \colhead{log($\xi$)} & \colhead{$N_H$} & \multicolumn{1}{c|}{$F_x$} & \colhead{log($\xi$)} & \colhead{$N_H$} & \colhead{$F_x$}
}
\startdata
xmm1 & $-0.67^{+0.05}_{-0.08}$ & $3.19^{+0.25}_{-0.44}$ & $< 5.21$ & $-0.67^{+0.05}_{-0.07}$ & $3.19^{+0.34}_{-0.43}$ & $< 5.16$ & $0.79^{+0.04}_{-0.03}$ & $7.97^{+0.59}_{-0.62}$ & unc. \\
xmm2 & $< -0.66$ & $2.02^{+0.29}_{-0.24}$ & unc. & $0.96^{+0.06}_{-0.09}$ & $4.76^{+0.66}_{-0.72}$ & unc. & $3.16^{+0.26}_{-0.29}$ & $4.06^{+1.22}_{-1.18}$ & unc. \\
xmm3 & $-0.27^{+0.06}_{-0.05}$ & $9.07^{+0.93}_{-2.43}$ & $2.66^{+0.90}_{-0.78}$ & $0.88^{+0.24}_{-0.10}$ & $8.69^{+0.22}_{-1.30}$ & $2.34^{+2.58}_{-1.12}$ & $2.57^{+0.44}_{-0.40}$ & $2.88^{+1.76}_{-1.44}$ & unc. \\
xmm4 & unc. & $0.55^{+0.70}_{-0.33}$ & unc. & unc. & unc. & unc. & $0.78^{+0.12}_{-0.05}$ & $7.67^{+1.00}_{-1.15}$ & unc. \\
xmm5 & $< -1.35$ & $2.13^{+0.19}_{-0.24}$ & $4.90^{+0.87}_{-0.67}$ & $0.79^{+0.04}_{-0.04}$ & $8.39^{+1.13}_{-0.61}$ & unc. & unc. & $< 3.37$ & unc. \\
xmm6 & $0.55^{+0.05}_{-0.10}$ & $> 6.04$ & unc. & $2.04^{+0.62}_{-0.16}$ & $0.99^{+1.03}_{-0.66}$ & unc. & $2.04^{+1.25}_{-0.16}$ & $0.91^{+0.81}_{-0.78}$ & unc. \\
xmm7 & $0.55^{+0.21}_{-0.17}$ & $> 5.27$ & unc. & $1.61^{+0.12}_{-0.39}$ & $1.50^{+1.69}_{-1.13}$ & unc. & $1.61^{+0.12}_{-0.41}$ & $1.50^{+2.80}_{-1.18}$ & unc. \\
xmm8 & $-0.07^{+0.05}_{-0.03}$ & $9.00^{+0.97}_{-0.92}$ & unc. & $1.63^{+0.12}_{-0.10}$ & $3.81^{+0.96}_{-0.92}$ & unc. & unc. & $1.96^{+1.29}_{-1.51}$ & unc. \\
xmm9 & $-0.05^{+0.19}_{-0.03}$ & $7.47^{+1.68}_{-1.26}$ & unc. & $1.63^{+0.15}_{-0.23}$ & $3.86^{+1.74}_{-0.99}$ & unc. & $> 1.64$ & $< 5.77$ & $0.66^{+0.38}_{-0.39}$ \\
\hline
xmmnu1 & - & - & unc. & - & - & unc. & - & - & $5.58^{+1.10}_{-0.64}$ \\
xmmnu2 & - & - & - & - & - & - & - & - & - \\
nu1 & $-0.80^{+0.17}_{-0.25}$ & $3.88^{+1.26}_{-0.77}$ & unc. & $0.79^{+0.06}_{-0.08}$ & $3.80^{+1.21}_{-1.61}$ & unc. & $1.63^{+0.01}_{-0.01}$ & $2.41^{+0.71}_{-0.67}$ & unc.
\enddata
\end{deluxetable*}

\section{Lag Measurements} \label{sec:rm}

\noindent We perform a spectral timing analysis using the software package \textsc{pyLag}\footnote{https://github.com/wilkinsdr/pyLag} to detect and measure X-ray reverberation from the inner accretion disk. Here, we briefly describe the methodology; a more comprehensive discussion is provided by \citet{Uttley2014}. The discrete Fourier transform (DFT) of an X-ray light curve, consisting of $N$ data points, evenly sampled with a cadence $\Delta t$, is given by
\begin{equation}
X_k = \sum_{n=1}^{N} x_n e^{-2\pi i kn/N}
\label{eq:dft}
\end{equation}
where $x_n$ is the $n$th data point of the light curve and $X_k$ represents the DFT at frequency $f_k = k/N\Delta t$. 

For two light curves $x(t)$ and $y(t)$, their DFTs $X_k$ and $Y_k$ can be expressed as 
\begin{align}
X_k & = |X_k| e^{i\phi_x} \\
Y_k & = |Y_k| e^{i\phi_y}
\label{eq:dft_phase}
\end{align}
where $|X_k|$ and $|Y_k|$ denote the amplitudes and $\phi_x$ and $\phi_y$ the phases. The cross spectrum $C_{XY,k}$ is defined as 
\begin{equation}
C_{XY,k} = X_k^* Y_k = |X_k| |Y_k| e^{i(\phi_y-\phi_x)}
\label{eq:crossspec}
\end{equation}
with $X_k^*$ representing the complex conjugate of $X_k$. The phase difference $\phi_y-\phi_x$ is directly related to the time lag between the light curves. 

To improve the signal-to-noise ratio (SNR) of the lag measurements, we compute the averaged cross spectrum $\overline{C}_{XY}(\nu_j)$ within frequency bins $\nu_j$ and over multiple light curve segments
\begin{equation}
\overline{C}_{XY}(\nu_j) = \frac{1}{KM} \sum_{k=i}^{i+K-1} \sum_{m=1}^{M} C_{XY,k,m}
\label{eq:crossspec_mean}
\end{equation}
where $\nu_j$ is the center of the $j$th frequency bin which comprises $K$ data points numbered from $i$ to $i+K-1$, and $M$ is the total number of light curve segments. The averaged time lag is then calculated as 
\begin{equation}
\tau(\nu_j) = \phi(\nu_j) / 2\pi \nu_j
\label{eq:lag}
\end{equation}
where $\phi(\nu_j)$ is the phase of the averaged cross spectrum $\overline{C}_{XY}(\nu_j)$. 

The coherence between the light curves is defined as 
\begin{equation}
\gamma^2(\nu_j) = \frac{|\overline{C}_{XY}(\nu_j)|^2 - n_p^2}{\overline{P}_X(\nu_j) \overline{P}_Y(\nu_j)}
\label{eq:coh}
\end{equation}
where $\overline{P}_X(\nu_j)$ and $\overline{P}_Y(\nu_j)$ are the power spectral densities (PSDs) in the frequency bin $\nu_j$ for the two light curves, and $n_p^2$ represents the contribution of Poisson noise to the amplitude of the cross spectrum. The coherence quantifies the fractional variability that is correlated between the light curves. It also characterizes the scatter of the cross spectra within the frequency bins and across the light curve segments due to non-coherent variability, so it provides an estimate of the uncertainty in the lag measurements:
\begin{align}
\sigma_\phi(\nu_j) & = \sqrt{\frac{1-\gamma^2(\nu_j)}{2\gamma^2(\nu_j)KM}} \\
\sigma_\tau(\nu_j) & = \sigma_\phi(\nu_j) / 2\pi \nu_j
\label{eq:lagerr}
\end{align}
where $\sigma_\phi(\nu_j)$ and $\sigma_\tau(\nu_j)$ are the uncertainties in the phase lag and time lag, respectively.

\subsection{Lag-Frequency Spectrum} \label{subsec:lagfreq}
\noindent We first compute the ``lag-frequency spectrum'', which represents the frequency-dependent time lag defined in Equation (\ref{eq:lag}), between a soft band 0.3 - 1 keV and an intermediate band 1.2 - 4 keV. This lag-frequency spectrum resolves the time lag in the low and high frequency Fourier components that describe the slow and fast components of the variability. We adopt ten logarithmically sampled frequency bins between $10^{-4}$ Hz and $10^{-3}$ Hz. To examine the overall frequency dependence of the lag, we stack the lag-frequency spectra of all \textit{XMM-Newton} epochs to increase the SNR. 

\begin{figure}
\epsscale{1.2}
\plotone{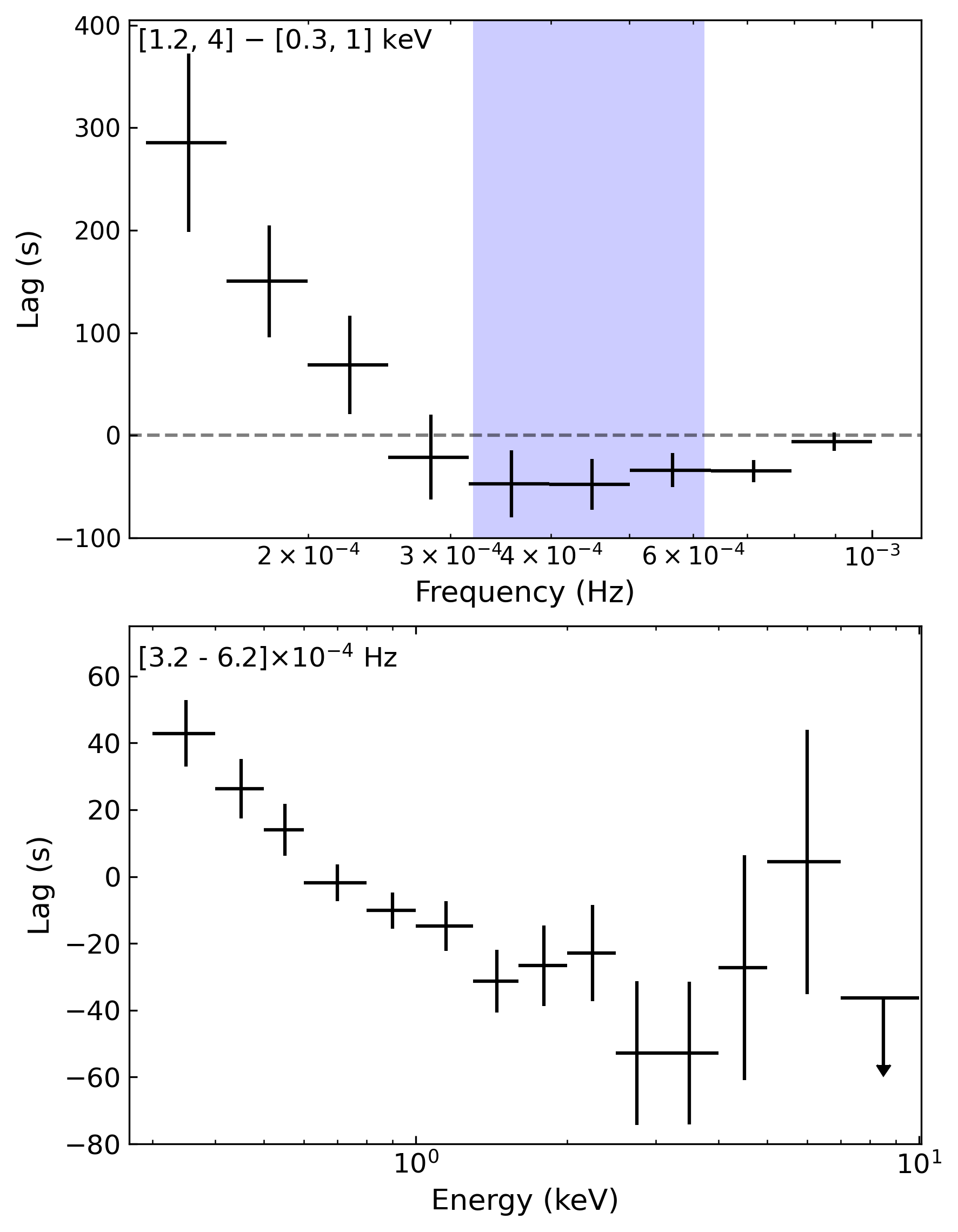}
\figcaption{\textit{(upper panel)} Stacked lag-frequency spectrum between the 1.2 - 4 keV band and the 0.3 - 1 keV band for all \textit{XMM-Newton} epochs. The lag is defined as positive when the hard-band variability lags behind the soft band. The shaded region covers the frequency range [3.2 - 6.2]$\times 10^{-4}$ Hz where the soft band lags behind the hard band due to light travel time from the corona to the disk. \textit{(lower panel)} Stacked lag-energy spectrum in [3.2 - 6.2]$\times 10^{-4}$ Hz for all \textit{XMM-Newton} epochs. The energy bin centered at 8.5 keV shows a lag of $\sim -300$ s with large uncertainty; for visibility, we display the $2\sigma$ upper limit of the lag in this bin. \label{fig:lagfreq}}
\end{figure}

Figure \ref{fig:lagfreq} (upper panel) shows the stacked lag-frequency spectrum. By definition, the high-frequency negative lags (highlighted by the shaded region) indicate that the soft band variability lags behind that of the intermediate band. Since the intermediate band is dominated by the continuum emission from the hot corona, whereas the soft band receives significant contributions from the disk reflection and the warm Compton component, the observed delay in the soft band at these frequencies can be attributed to the light travel time from the corona to the disk.

In contrast, the low-frequency positive lags indicate that the intermediate band variability lags behind that of the soft band, a phenomenon that cannot be explained by the disk reflection scenario. The origin of the low-frequency lag is not fully understood, with potential explanations including the inward propagation of the fluctuations originating in the outer regions of the accretion disk or corona \citep[e.g.,][]{Kotov2001,Arevalo2006}. Given the primary focus of this paper is on the X-ray reverberation driven by disk reflection, our analysis concentrates on the high-frequency lags. 

\subsection{Lag-Energy Spectrum} \label{subsec:lagen}
\noindent To investigate the energy dependence of the lag, we generate light curves in 14 energy bins covering 0.3 - 10 keV. For each energy bin, we compute the time lag between the variability in that bin and the average of all other bins over the frequency range [3.2 - 6.2]$\times 10^{-4}$ Hz (the shaded region in the upper panel of Figure \ref{fig:lagfreq}). The lower panel of Figure \ref{fig:lagfreq} shows the stacked lag-energy spectrum for all \textit{XMM-Newton} epochs. The highest energy bin exhibits a lag of $\sim -300$ s with large uncertainty due to the low count rate; for visibility, we display the $2\sigma$ upper limit for this bin. Except for the highest energy bin, the photons in the $\sim$ 2.5 - 4 keV range have the earliest arrival times, while the soft photons and, to a less significant extent, the $\sim$ 5 - 7 keV photons, are relatively delayed. 

Since the lag-energy spectrum is computed relative to the mean of all other bins, the absolute response time in each individual bin is arbitrary. The shape of the lag-energy spectrum can be explained by the intrinsic energy dependence of the disk reflection lag combined with a dilution effect. Each energy bin contains contributions from both the coronal continuum and disk reflection. Since the continuum has zero lag relative to itself, its contribution ``dilutes'' the intrinsic lag signal of the disk reflection. Consequently, energy bins with a larger fraction of continuum show smaller observed lags, whereas the soft band and the 5 - 7 keV band, which are more strongly influenced by the soft excess and the Fe K$\alpha$ line respectively, exhibit larger lags \citep[e.g.,][]{Wilkins2013,Cackett2014}.  

\begin{figure*}
\gridline{\fig{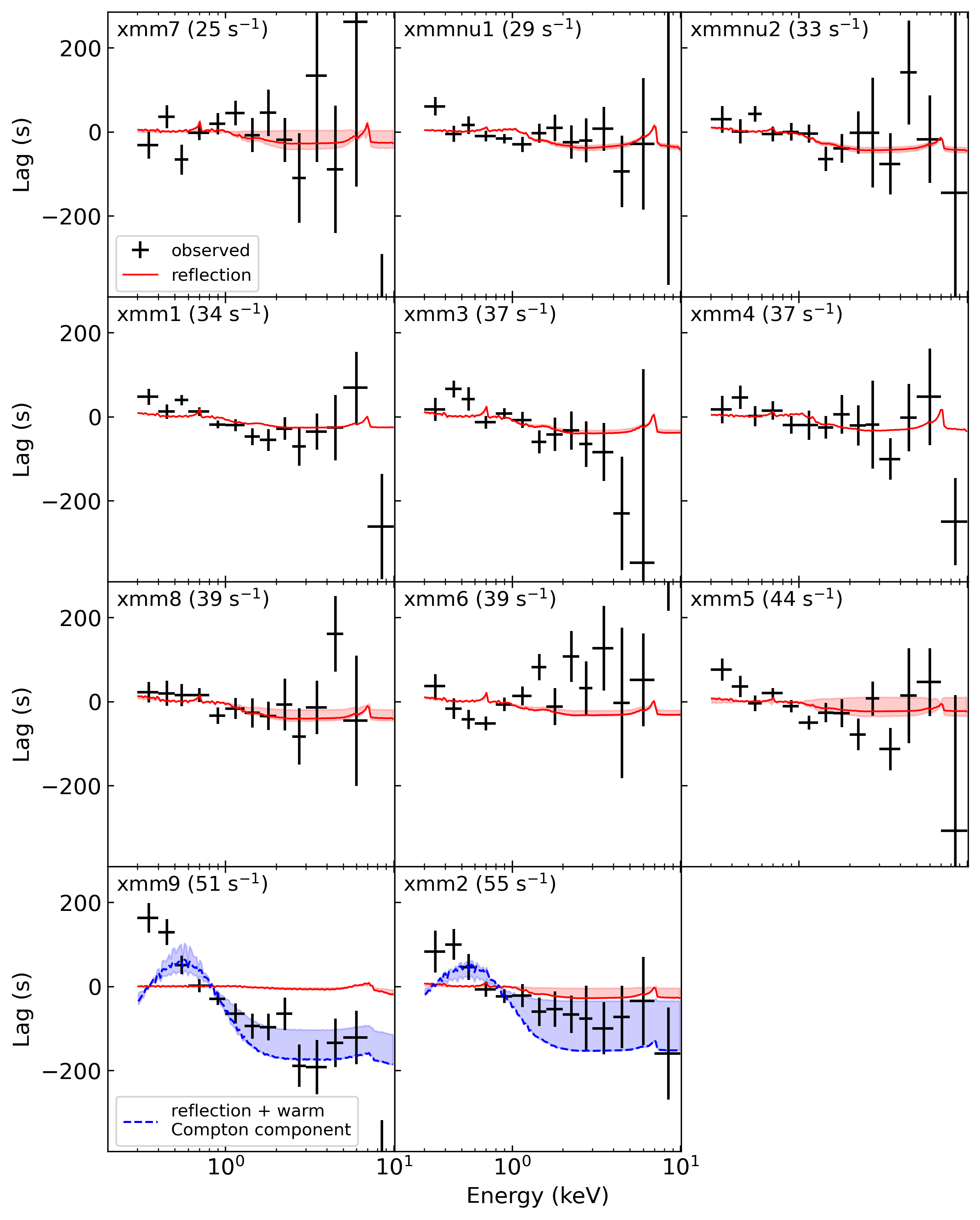}{0.93\textwidth}{}}
\figcaption{Lag-energy spectrum in the frequency range [3.2 - 6.2]$\times 10^{-4}$ Hz. Each panel corresponds to an \textit{XMM-Newton} epoch, with its name and count rate given in the upper left corner. The panels are arranged in the order of increasing count rate. The black crosses show the observed lag-energy spectra. The colored lines represent the model lag-energy spectra based on the spectral parameters and ray-tracing simulations (see Section \ref{sec:sim}). We do not fit the models to the observed lag-energy spectra. The red solid lines represent models in which the reverberation signal arises only from disk reflection, while the blue dashed lines in the bottom two panels represent models that also include reverberation from the warm Compton component. The shaded regions denote the 1$\sigma$ uncertainty in the models due to the uncertainty in the corona height $h$ from the spectral fitting. For xmm1 and xmm4 where the corona height is poorly constrained, we generate models based on the $2\sigma$ lower or upper limit of the corona height without providing uncertainty estimates. \label{fig:lagen}}
\end{figure*}

\begin{figure}
\epsscale{1.2}
\plotone{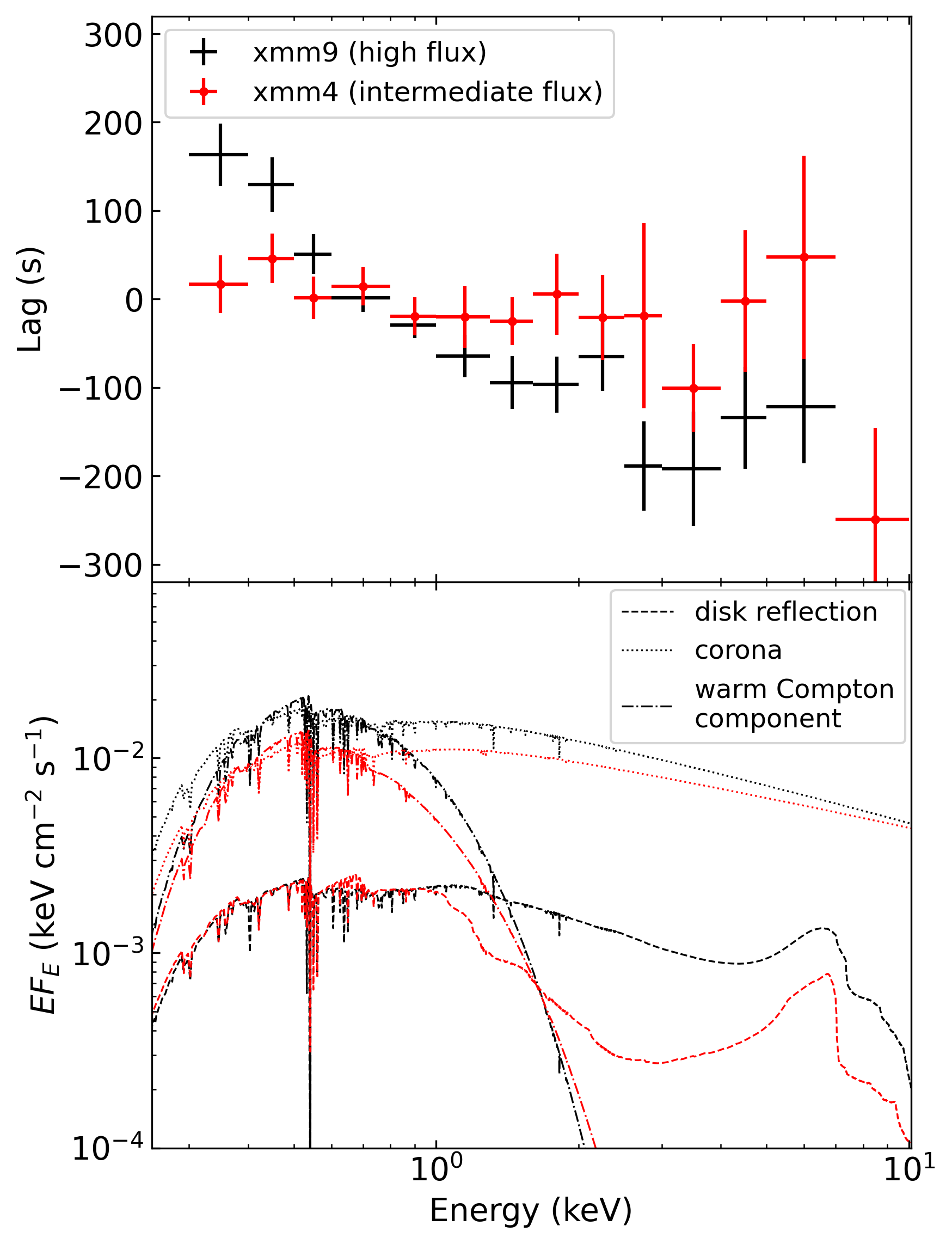}
\figcaption{\textit{(upper panel)} Comparison of the lag-energy spectra of a high-flux epoch xmm9 (black) and an intermediate-flux epoch xmm4 (red). Since the lag in each bin is computed relative to the mean of all other energy bins, the zero point of the lag-energy spectrum is not critical. \textit{(lower panel)} Comparison of the best-fit \textsc{XSPEC} models for xmm9 (black) and xmm4 (red). The dashed, dotted, and dash-dotted lines represent the disk reflection \textsc{relxilllpCp}, the coronal continuum \textsc{nthcomp}, and the warm Compton component \textsc{comptt}, respectively. The transmitted emission from the warm absorbers contributes negligibly to the total flux and broadband spectral shape, so we do not show them in this figure for visibility. \label{fig:lagen_spcomp}}
\end{figure}

We observe significant variation in the lag-energy spectrum between low and high flux epochs. Figure \ref{fig:lagen} shows the lag-energy spectra for all individual epochs sorted by increasing count rate, while Figure \ref{fig:lagen_spcomp} (upper panel) compares the lag-energy spectra between an intermediate-flux epoch xmm4 and a high-flux epoch xmm9. The low-flux epochs exhibit relatively flat lag-energy spectra, lacking significant soft excess or Fe K$\alpha$ line features. As the X-ray flux increases, an excess of soft lag becomes apparent and grows. In particular, high-flux epochs such as xmm9 and xmm2 exhibit strong soft excesses that differ significantly from the flat lag-energy spectra observed in low-flux epochs. However, not all epochs follow this trend. For example, the lag-energy spectrum of xmm6 shows hints of a dip in the soft bands that is not observed in other epochs. Minor modifications to the frequency range used to derive the lag-energy spectra do not alter the qualitative evolution of the lags. Although the reverberation lags in the low-flux epochs xmm7 and xmmnu1 peak at slightly higher Fourier frequencies than in the high-flux epochs, their peak amplitudes remain smaller than in the high-flux epochs. 

The Fe K$\alpha$ feature in the lag-energy spectra exhibits more complex behavior. Low-flux epochs do not display clear Fe K$\alpha$ lags, which may either be due to an intrinsic absence of Fe K$\alpha$ reverberation or due to large uncertainties in the high-energy bins. The Fe K$\alpha$ feature is more pronounced in some intermediate-flux epochs, such as xmm1, xmm4, and xmm5, while it is not significantly detected in others. High-flux epochs show hints of the Fe K$\alpha$ line, but the feature is weaker compared to the that observed in intermediate-flux epochs. 

Figure \ref{fig:lagen_spcomp} (lower panel) compares the best-fit \textsc{XSPEC} models between xmm4 and xmm9. Both epochs have small reflection fractions, meaning that the disk reflection component is weak relative to the coronal continuum. The large dilution factor resulting from the small reflection fraction may contribute to the weak Fe K$\alpha$ features observed in the lag-energy spectra. Nevertheless, the lag-energy spectrum of the high-flux epoch xmm9 exhibits a substantial soft lag, despite having an even lower reflection fraction than the intermediate-flux epoch xmm4.

\section{Simulations and Lag Modeling} \label{sec:sim}
\noindent To better understand the evolution of the spectral and timing properties, we create a reverberation lag-energy model for each epoch, based upon the parameters inferred from modeling the X-ray flux spectrum (in Section \ref{sec:spec}) and a general relativistic ray-tracing simulation \citep{Wilkins2012,Wilkins2013,Wilkins2016}. We assume a black hole with spin parameter $a$, a point-source lamppost corona located at a height $h$ above the black hole, and a standard geometrically thin, optically thick accretion disk \citep{Shakura1973}. The simulation traces the propagation of rays in a Kerr spacetime from the corona to the disk and then reflected from the disk to a distant observer. By counting the number of photons received as a function of energy $E$ and arrival time $t$, the simulation predicts an impulse response function $\Psi(E/E_0,t)$. This response function characterizes the response of the disk seen by the observer to a $\delta$-function flare in the corona at an energy $E_0 = 6.4$ keV. 

We construct a grid of response functions $\Psi(E/E_0,t)$ with 23 values of corona height $h$ ranging from 1.5$r_g$ to 50$r_g$, 12 values of black hole spin $a$ from 0 to 0.998, and 16 values of disk inclination from 5 to 80 degrees. For each epoch, we select the response function whose parameters $h$, $a$ and inclination are closest to the best-fit spectral parameters listed in Table \ref{tab:specpar}. We then convolve the response function $\Psi(E/E_0,t)$ with a spectral model $s_r(E)$,
\begin{equation}
S_r(E,t) = s_r(E) \circledast \Psi(E/E_0,t)
\label{eq:entcov}
\end{equation}
adopting the non-relativistic reflection model \textsc{xillver} for $s_r(E)$. In this model, the continuum slope $\Gamma$, iron abundance $A_{\rm Fe}$, ionization parameter log($\xi$), and disk inclination are set to the best-fit values in Table \ref{tab:specpar}. The general relativistic blurring due to Doppler shifts and gravitational redshifts, in addition to the light travel time, is encoded by the response function $\Psi(E/E_0,t)$, so $S_r(E,t)$ becomes the relativistic reflection model used in our spectral modeling when integrated over the arrival time $t$. 

We then add the primary continuum to the response function, to self-consistently incorporate dilution into the model. We define a continuum response function 
\begin{equation}
S_p(E,t)=\begin{cases}
    E^{-\Gamma}, & \text{if $t=t_0$}.\\
    0, & \text{otherwise}.
  \end{cases}
\label{eq:entcont}
\end{equation}
where $\Gamma$ is the power-law slope of the continuum and $t_0$ is the arrival time of photons coming directly from the corona without being reflected by the disk. We add the continuum response function to the convolved disk response to obtain
\begin{equation}
S(E,t) = A_r S_r(E,t) + A_p S_p(E,t)
\label{eq:ent_addcont}
\end{equation}
The normalizations $A_r$ and $A_p$ are determined such that 
\begin{equation}
\frac{A_r \iint S_r(E,t) dt\,dE}{A_p \iint S_p(E,t) dt\,dE} = \frac{F_r}{F_p}
\label{eq:ent_reffrac}
\end{equation}
where $F_r$ and $F_p$ denote the integrated photon fluxes of the reflection model \textsc{relxilllpCp} and the continuum model \textsc{nthcomp}, respectively, with the integration performed over the energy range 0.3 - 10 keV for both sides of the equation. 

To compare with the observed lag-energy spectrum, we need to account for the frequency dependence of the lags. The total response $S(E,t)$ is equivalent to a set of model light curves as a function of energy $E$ following a $\delta$-function flare in the corona. We therefore process the response function $S(E,t)$ using the same Fourier analysis methodology as described in Section \ref{sec:rm} to compute the model lag-energy spectrum over the same frequency range as the observations. We then convert the time units from $GM_{\rm BH}/c^3$ used in the model calculations to absolute units adopting a single-epoch SMBH mass estimate of ${\rm log}(M_{\rm BH}/M_{\odot})=6.27$ for Ark 564 \citep{Zhang2006}.

The corona height $h$ is a key parameter in determining the lags but is subject to large uncertainties from the spectral fitting of most epochs. We use a Monte-Carlo method to estimate the uncertainty in the model lag-energy spectrum due to $h$. Specifically, we draw 500 realizations of $h$ from a Gaussian distribution centered at the best-fit value with a standard deviation $\sigma_{\rm h, up}$ and another 500 realizations from a Gaussian with the same center but a standard deviation $\sigma_{\rm h, low}$, where $\sigma_{\rm h, up}$ and $\sigma_{\rm h, low}$ represent the upper and lower uncertainties of $h$. We then discard the realizations in the first distribution that fall below the best-fit value and those from the second distribution that exceed the best-fit. For each remaining realization, we compute a model lag-energy spectrum using the procedure described above. The scatter among these realizations provides an estimate of the uncertainty in the model lag-energy spectra. 

Figure \ref{fig:lagen} shows the model lag-energy spectra generated by this procedure (red solid lines), with the red shaded region representing the 1$\sigma$ uncertainty of the model due to the corona height uncertainty. For xmm1 and xmm4 where the corona height is poorly constrained, we generate models based on the $2\sigma$ lower or upper limit of the corona height without providing uncertainty estimates. Despite the unusual features in observed lag-energy spectra of xmm3 and xmm6, the model lag-energy spectra based on the disk reflection model broadly agree with the observations for the low-flux epochs up to xmm5. The observed lag-energy spectra of xmm5 begin to exhibit an excess of the soft lags relative to the model. In the high flux epochs xmm9 and xmm2, where the soft excess is found to lag significantly behind the primary continuum, the simple reverberation model (a point source corona illuminating a flat accretion disk) fails to reproduce the observed lag - energy spectra. The weak Fe K$\alpha$ lags predicted by the model, due to small reflection fractions, agree with the observed weakness or absence of the Fe K$\alpha$ feature in most epochs. However, the observed lag-energy spectra of xmm1 and xmm5 show stronger Fe K$\alpha$ lags than predicted by the model, although these can still marginally match within the large uncertainties.

\subsection{Reverberation Response of the Soft Excess} \label{subsec:simcpt}
\noindent We test whether including the soft excess, modeled by the Comptonization of disk photons in a warm plasma above the disk, in the reverberation model is able to reproduce the observed time lags in high flux epochs. As a starting point, we assume that the warm Compton component is a thin layer of atmosphere on the disk surface, covering the same radial range as the disk reflection model and characterized by a uniform plasma temperature, seed photon temperature and optical depth across its extent. Variations in the coronal continuum flux alter the heating of the accretion disk, which in turn modifies the production of UV photons. This change affects the number of seed photons available for Comptonization and, consequently, the normalization of the warm Compton emission. Under this scenario, the warm Compton component responds to the corona variability with the same time delay as the disk reflection.

To implement this scenario in our model, we introduce an additional term to Equation (\ref{eq:ent_addcont}), yielding 
\begin{equation}
S(E,t) = [A_r s_r(E) + A_c s_c(E)]\circledast \Psi(E/E_0,t) + A_p S_p(E,t)
\label{eq:ent_addcpt}
\end{equation}
where $s_c(E)$ is the \textsc{comptt} model representing the warm Compton component, with parameters set according to the spectral fitting results in Table \ref{tab:specpar}. We use the same response function $\Psi(E/E_0,t)$ for the disk reflection and the soft excess modeled by warm Compton component, which describes a warm atmosphere extending across the surface of the disk. The normalization $A_c$ is determined such that 
\begin{equation}
\frac{A_r \int s_r(E) dE}{A_c \int s_c(E) dE} = \frac{F_r}{F_c}
\label{eq:ent_Rcpt}
\end{equation}
where $F_r$ and $F_c$ denote the integrated photon flux of the reflection model \textsc{relxilllpCp} and the warm Compton component \textsc{comptt}, respectively, over the energy range 0.3 - 10 keV. We then derive the model lag-energy spectra following the same procedure as the disk-reflection-only model. The blue dashed lines in the bottom panels of Figure \ref{fig:lagen} show the model lag-energy spectra after including the reverberation of the warm Compton component. It provides significantly better agreement with the observations in the high-flux epochs xmm9 and xmm2 compared to the disk-reflection-only model.

\subsection{Temperature Gradient in the Comptonizing Atmosphere} \label{subsec:simcr}
\noindent While the model presented in Section \ref{subsec:simcpt} better matches the observations of the high-flux epochs, the soft end of the observed lag-energy spectra for xmm9 remains higher than predicted. To reconcile this discrepancy, we extend the reverberation model of the warm Compton component by incorporating the radial temperature gradient of the disk. This gradient is expected in the standard accretion disk theory \citep{Shakura1973} and causes the seed photons that feed the warm Compton component to have different energies at different radii, leading to a radial dependence of the warm Compton emission. To account for this, we modify Equation (\ref{eq:ent_addcpt}) as follows: 
\begin{multline}
S(E,t) = A_c' \int_r s_c[E,T_s(r)] \circledast \psi(E/E_0,t,r)dr
\\+ A_r s_r(E) \circledast \Psi(E/E_0,t) + A_p S_p(E,t)
\label{eq:ent_cr}
\end{multline}
where $s_c[E,T_s(r)]$ is the \textsc{comptt} model for the warm Compton component with a seed photon temperature $T_s(r)$, and $\psi(E/E_0,t,r)$ is the response function at radius $r$ produced by the simulation. We determine the normalization $A_c'$ by requiring that the integrated flux of the temperature-gradient \textsc{comptt} model equals that of the constant-temperature model, i.e.,
\begin{multline}
A_c' \int_E \int_r s_c[E,T_s(r)] \varepsilon(r)rdrdE
\\ = A_c \int_E s_c[E,\overline{T_s}] dE \int_r  \varepsilon(r)rdr
\label{eq:ent_crnorm}
\end{multline}
where $\varepsilon(r)$ is the disk emissivity profile for the lamppost geometry from the simulation, and $\overline{T_s} = 0.1$ keV is the seed photon temperature used in the constant-temperature \textsc{comptt} model. 

\begin{figure}
\epsscale{1.2}
\plotone{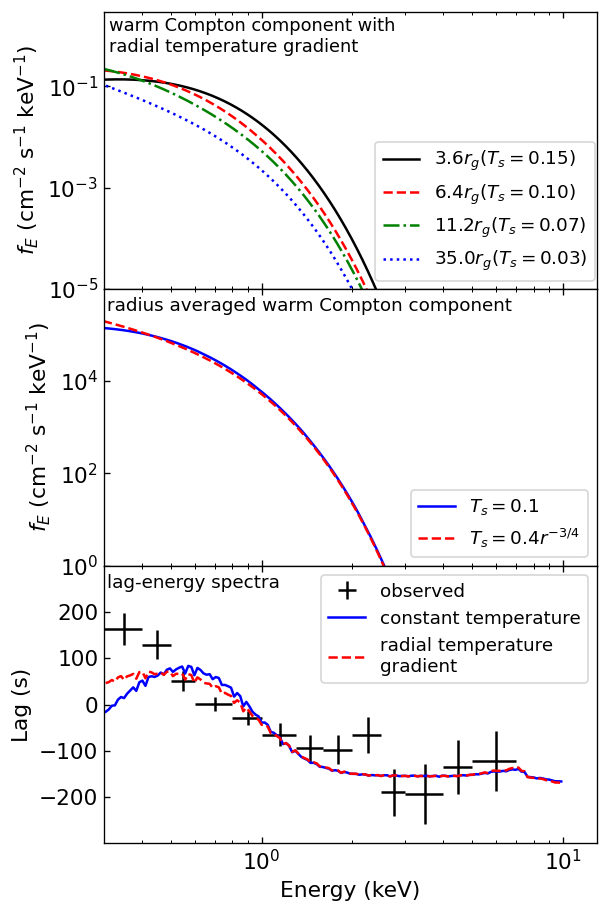}
\figcaption{\textit{(top panel)} Warm Compton component $s_c[E,T_s(r)]$ with different seed photon temperatures. The black solid, red dashed, green dash-dotted and blue dotted lines correspond to models with seed photon temperatures of 0.15, 0.1, 0.07 and 0.03 keV, respectively. These temperatures correspond to radii of $3.6r_g$, $6.4r_g$, $11.2r_g$ and $35r_g$ for a temperature profile $T_s(r) = 0.4\, r^{-3/4}$ keV. \textit{(middle panel)} Comparison of the temperature-gradient warm Compton component model $s_c[E,T_s(r)]$ averaged over radius (red dashed line) with the constant-temperature model $s_c[E,\overline{T_s}]$ (blue solid line). \textit{(bottom panel)} Lag-energy spectra for xmm9. The black crosses represent the observed lag-energy spectrum, while the blue solid and red dashed lines represent the model lag-energy spectra based on the constant-temperature and temperature-gradient warm Compton models, respectively. \label{fig:lagen_cr}}
\end{figure}

We adopt a temperature profile of the standard thin disk given by
\begin{equation}
T_s(r) = T_{s0} \, r^{-3/4}
\label{eq:Tsr}
\end{equation}
The top panel of Figure \ref{fig:lagen_cr} shows the model of the warm Compton component $s_c[E,T_s(r)]$ at different radii based on this profile with $T_{s0} = 0.4$ keV. To minimize the difference between the emissivity-weighted average of $s_c[E,T_s(r)]$ over radius with the constant-temperature model $s_c[E,\overline{T_s}]$ (middle panel of Figure \ref{fig:lagen_cr}), we tested multiple values of $T_{s0}$ between 0.1 and 0.4 keV. We find the emissivity-weighted average of $s_c[E,T_s(r)]$ most closely matches the constant-temperature model when $T_{s0} = 0.4$ keV. Larger values of $T_{s0}$ would cause the seed photon temperature at the innermost radius to exceed the plasma temperature of the warm Compton component, thereby violating the Compton heating scenario. Therefore, we adopt $T_{s0} = 0.4$ keV for our fiducial temperature profile. 

The bottom panel of Figure \ref{fig:lagen_cr} shows the predicted lag-energy spectrum from the combined reflection plus temperature-gradient warm Compton component. The soft end of this model is elevated compared to the constant-temperature model described in Section \ref{subsec:simcpt}. This is because the soft photons in the temperature-gradient model preferentially originate from the warm Comptonizing plasma at larger radii, where cooler seed photons are present. The longer light travel time to the outer disk results in increased soft lags. However, the soft end in the temperature-gradient model still falls short of fully matching the observed lag-energy spectrum. A more extreme temperature gradient that produces an even cooler outer disk may further elevate the soft end of the lag–energy spectrum. Likewise, adopting a flatter emissivity profile for the warm atmosphere that enhances emission from large radii would also tend to increase the soft lags. Nevertheless, implementing these modifications risks over-complicating our model without strong physical motivation, so we do not explore them in detail.

\subsection{Reverberation Mechanism within the Comptonizing Atmosphere} \label{subsec:simthpl}
\noindent We examine mechanisms by which continuum variations can reverberate in the Comptonizing atmosphere. In addition to the modulation of seed photon flux (by disk heating) discussed in Sections \ref{subsec:simcpt} and \ref{subsec:simcr}, another mechanism is the direct reflection of the coronal continuum by the warm Comptonizing atmosphere. In this context, the ``reflection'' refers to the process where a fraction of coronal photons are Compton-scattered within the warm plasma before escaping to infinity, resulting a Comptonized power-law spectrum. Under the same assumption as in Section \ref{subsec:simcpt}, where the warm atmosphere is a thin layer on the disk surface, the warm atmosphere reflection responds to continuum variability after a light travel delay from the corona to the disk. 

\begin{figure}
\epsscale{1.2}
\plotone{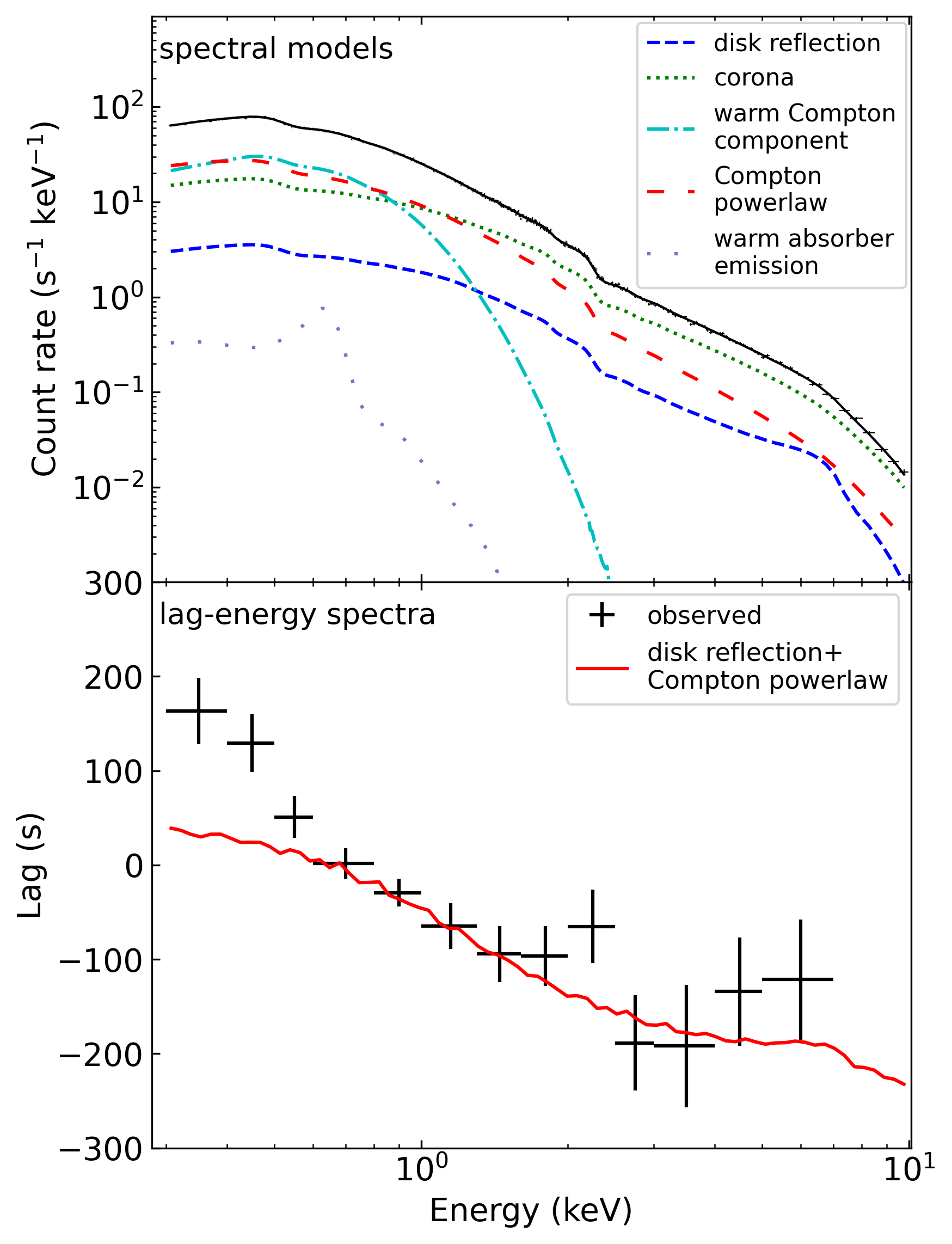}
\figcaption{\textit{(upper panel)} X-ray spectra and models for xmm9 after incorporating the Comptonized power-law \textsc{thcomp*powerlaw} to model warm atmosphere reflection. The black crosses denote the observed spectra, and the black solid line indicates the best-fit spectral model. The blue dashed, green dotted, cyan dash-dotted, red sparsely dashed and purple sparsely dotted lines represent the disk reflection \textsc{relxilllpCp}, the coronal continuum \textsc{nthcomp}, the warm Compton component \textsc{comptt}, the warm atmosphere reflection \textsc{thcomp*powerlaw}, and the transmitted emission from the warm absorbers, respectively. \textit{(lower panel)} Lag-energy spectra for xmm9. The black crosses show the observed lag-energy spectrum, while the red solid line represents the prediction of the combined disk reflection plus warm atmosphere reflection model. \label{fig:lagen_thpl}}
\end{figure}

We model this warm atmosphere reflection using a Comptonized power-law component \textsc{thcomp*powerlaw} and incorporate it into the joint spectral fitting of all epochs in \textsc{XSPEC}. The power-law index is tied to that of the coronal continuum model \textsc{nthcomp}, and the plasma temperature is tied to that of the \textsc{comptt} component modeling the Comptonization of disk photons by the warm atmosphere. Because the trajectories of the coronal photons reflected by the warm atmosphere are likely different from those of the disk photons being Comptonized, we allow the optical depth of \textsc{thcomp} to vary freely. For the epoch xmm9, we obtain an optical depth of $\sim 35$ for the \textsc{thcomp} model, while the best-fit plasma temperature $kT \approx 0.13$ keV is similar to the fiducial model in Table \ref{tab:specpar}. The upper panel of Figure \ref{fig:lagen_thpl} shows the spectral models for xmm9 after including the \textsc{thcomp*powerlaw} component. The spectral shape of the Comptonized power-law \textsc{thcomp*powerlaw} is close to a power-law, while it is softer than the coronal continuum due to the effects of Comptonization in the warm plasma. 
 
We then examine whether including warm atmosphere reflection can yield a better model for the lag-energy spectrum of xmm9. The bulk of the \textsc{comptt} component in the average spectrum is the direct Comptonization of UV seed photons, which is slowly varying and linked to the variability timescales within the disk. This variation of the \textsc{comptt} component does not reverberate in response to variations in the corona. In the reverberation model, we consider the smaller, short-timescale variability on top of this, driven by coronal continuum variations, reverberating off the disk and its atmosphere. Accordingly, we modify Equation (\ref{eq:ent_addcont}) as follows:
\begin{equation}
S(E,t) = [A_r s_r(E) + A_t s_t(E)]\circledast \Psi(E/E_0,t) + A_p S_p(E,t)
\label{eq:ent_thpl}
\end{equation}
where $s_t(E)$ represents the warm atmosphere reflection modeled by the Comptonized power-law \textsc{thcomp*powerlaw}. We then compute the model lag-energy spectra using the same methodology as in Section \ref{subsec:simcpt}. 

The lower panel of Figure \ref{fig:lagen_thpl} shows the lag-energy spectra predicted by the combined disk reflection plus warm atmosphere reflection model. The model broadly agrees with the observations down to $\sim 0.5$ keV, but the soft end of the model remains too flat to fully match the observed lag-energy spectrum. Overall, this direct reverberation model of continuum photons in the warm atmosphere provides comparable results to the the modulation of \textsc{comptt} flux discussed in Sections \ref{subsec:simcpt} and \ref{subsec:simcr}, although it is slightly less consistent with the observations at the hard end.

\section{Discussion} \label{sec:discussion}

\subsection{Evolution of Lags} \label{subsec:lagtrend}
\begin{figure*}
\gridline{\fig{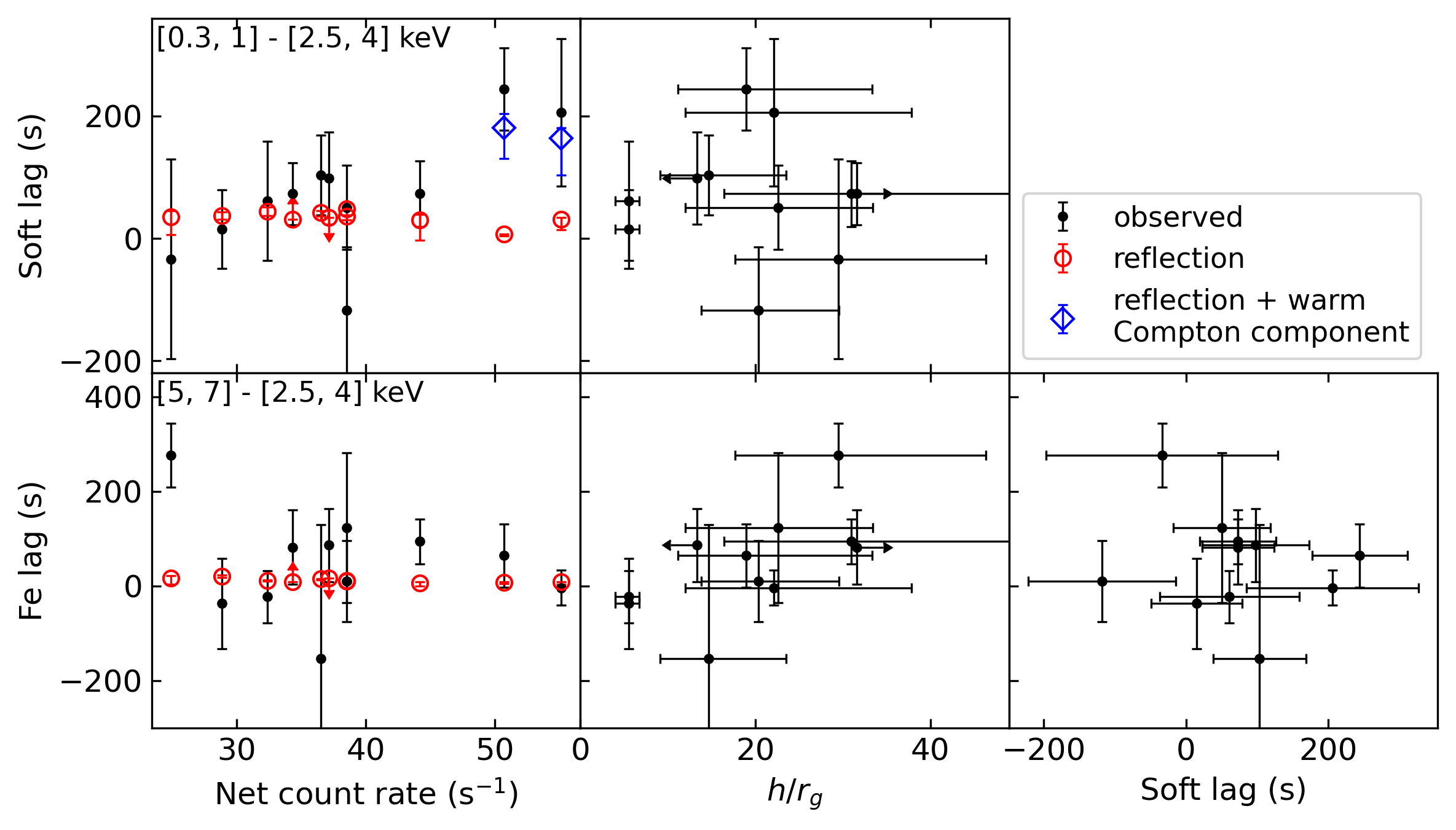}{\textwidth}{}}
\figcaption{Soft lag (upper row) and Fe K$\alpha$ lag (lower row) as functions of net count rate (left column), corona height (middle column), and soft lag (right column). The soft lag is defined as the time delay between the 0.3 - 1 keV and 2.5 - 4 keV bands, whereas the Fe K$\alpha$ lag is defined as the time delay between the 5 - 7 keV and 2.5 - 4 keV bands. A lag is considered positive when the variability in the first band lags behind that in the second band. All lags are calculated within a frequency range of [3.2 - 6.2]$\times 10^{-4}$ Hz. The red circles in the left column represent the lags predicted by the reverberation model based on disk reflection (Section \ref{sec:sim}), while the blue diamonds represent the lags predicted by the combined disk reflection plus warm Compton component (Section \ref{subsec:simcpt}). The uncertainties in the model lags are derived using the same Monte Carlo method applied to estimate the uncertainties of the model lag–energy spectra, which are driven by the corona height uncertainties (Section \ref{sec:sim}). \label{fig:lagtrend}}
\end{figure*}

\noindent To investigate the correlation between reverberation lags and other spectral and timing properties, we define two types of broad-band lags: a soft lag between the 0.3 - 1 keV and 2.5 - 4 keV bands, and a Fe K$\alpha$ lag between the 5 - 7 keV and 2.5 - 4 keV bands. We compute these lags for all \textit{XMM-Newton} epochs within the same frequency range [3.2 - 6.2]$\times 10^{-4}$ Hz used for the lag-energy spectra. Figure \ref{fig:lagtrend} (left column) shows the broad-band lags as a function of net count rate. The soft lag increases with count rate, consistent with the evolution observed in the lag-energy spectra in Figure \ref{fig:lagen}. This trend is relatively weak in low-flux epochs due to large uncertainties but becomes significant in the two high-flux epochs. In contrast, the Fe K$\alpha$ lags do not show a clear correlation with count rate. We only find Fe K$\alpha$ line features in the lag-energy spectra for xmm1, xmm4, xmm5, and potentially xmm9, while the apparent Fe K$\alpha$ lags in other epochs are likely dominated by statistical fluctuations. 

The red circles and blue diamonds in the left column of Figure \ref{fig:lagtrend} represent the predicted lag from the models described in Section \ref{sec:sim}. The soft lags predicted by the disk-reflection-only model broadly agree with the observations except in the two high-flux epochs. The disk reflection + warm Compton component model predicts soft lags that are consistent with the observations within the 1$\sigma$ range, although its model lag-energy spectrum does not fully match the observation for xmm9 in the two softest bins. While the model Fe K$\alpha$ lags are small due to the dilution effect, they are marginally consistent with the observations except for xmm7 and xmm5. The apparent discrepancy between the large observed Fe K$\alpha$ lag and the model in the lowest-flux epoch xmm7 is likely attributable to statistical fluctuations, as no robust Fe K$\alpha$ line feature is detected in its lag-energy spectrum. 

One potential cause for the observed lag variability is a change in the corona height. The corona height influences the light travel time from the corona to the disk, thereby affecting the reverberation lags. The middle column of Figure \ref{fig:lagtrend} shows the broad-band lags as a function of the corona height derived from spectral fitting. There is no clear correlation between the soft lag and the corona height. Although the Fe K$\alpha$ lag exhibits a tentative increase with larger corona heights, this correlation is not robust because the Fe K$\alpha$ line is not detected in the lag-energy spectra for most epochs. The absence of a correlation may suggest that the corona height is not the primary driver of lag variability. However, it is also possible that the intrinsic correlation is obscured by large uncertainties. 

An alternative explanation for the soft lag variability is the reverberation of the warm Compton component. For the high-flux epochs xmm9 and xmm2, the predicted lags from the combined disk reflection plus warm Compton component reverberation model agree reasonably well with the observations, while the disk-reflection-only model significantly under-predicts the lag. This indicates that the excess soft lag in the high-flux epochs is likely due to contributions from the warm Compton component. The reverberation of the warm Compton component also explains the lack of correlation between the soft lag and the Fe K$\alpha$ lag shown in the right column of Figure \ref{fig:lagtrend}, since the warm Compton component contributes predominantly to the soft bands. 

Our spectral and timing results support the presence of a warm Comptonized layer on the disk surface that contributes to the soft excess. Radiative‐transfer studies have shown that such a warm atmosphere with a temperature of $kT_e \sim 0.1 - 1$ keV and an optical depth of $\tau_p \sim 20$ can form when illuminated by the coronal continuum from above and the disk black-body emission from below \citep[e.g.,][]{Ballantyne2020,Petrucci2020}. They also predict that, under appropriate conditions, this warm atmosphere will not exhibit strong emission or absorption lines. These predictions are consistent with our requirement for a warm Compton component in spectral fitting and with the best-fit parameters we obtain. Furthermore, \citet{Petrucci2020} discuss the potential contribution of the warm atmosphere to reverberation lags, which we observe in the lag-energy spectra of the high-flux epochs.

\subsection{Covariance Spectra} \label{subsec:cov}
\noindent While both the coronal continuum and disk reflection are variable, their variability is not necessarily fully correlated. The variability of the warm Compton component and its potential correlation with the continuum and disk reflection are also uncertain. In principle, only the covariance, which characterizes the correlated component of the variability, contributes to the observed lag signals. In the reverberation modeling described in Section \ref{sec:sim}, we employ spectral models derived from \textsc{XSPEC} fitting of the mean spectra, which implicitly assumes that the covariance spectra have the same shape as mean spectra. In this section, we compute the covariance spectra and compare them with the mean spectra to assess the validity of this assumption. 

The covariance between two light curves $x(t)$ and $y(t)$ is given by
\begin{equation}
Cv(\nu_j) = \langle x \rangle \sqrt{\frac{\Delta\nu_j (|\overline{C}_{XY}(\nu_j)|^2 - n_p^2)}{\overline{P}_Y(\nu_j) - P_{Y,\rm noise}}} 
\label{eq:cov}
\end{equation}
where $\Delta\nu_j$ is the width of the frequency bin $\nu_j$, $\overline{C}_{XY}(\nu_j)$ is the cross spectrum, $n_p$ is the Poisson noise contribution to the cross spectrum amplitude, $\overline{P}_Y(\nu_j)$ is the PSD of $y(t)$, and $P_{Y,\rm noise}$ is the contribution of Poisson noise to the observed PSD. To derive the covariance spectrum, we adopt the same energy binning as for the lag-energy spectrum and compute the covariance of each energy bin relative to the sum of all the other energy bins, within a frequency range [3.2 - 6.2]$\times 10^{-4}$ Hz. 

\begin{figure}[ht]
\epsscale{1.2}
\plotone{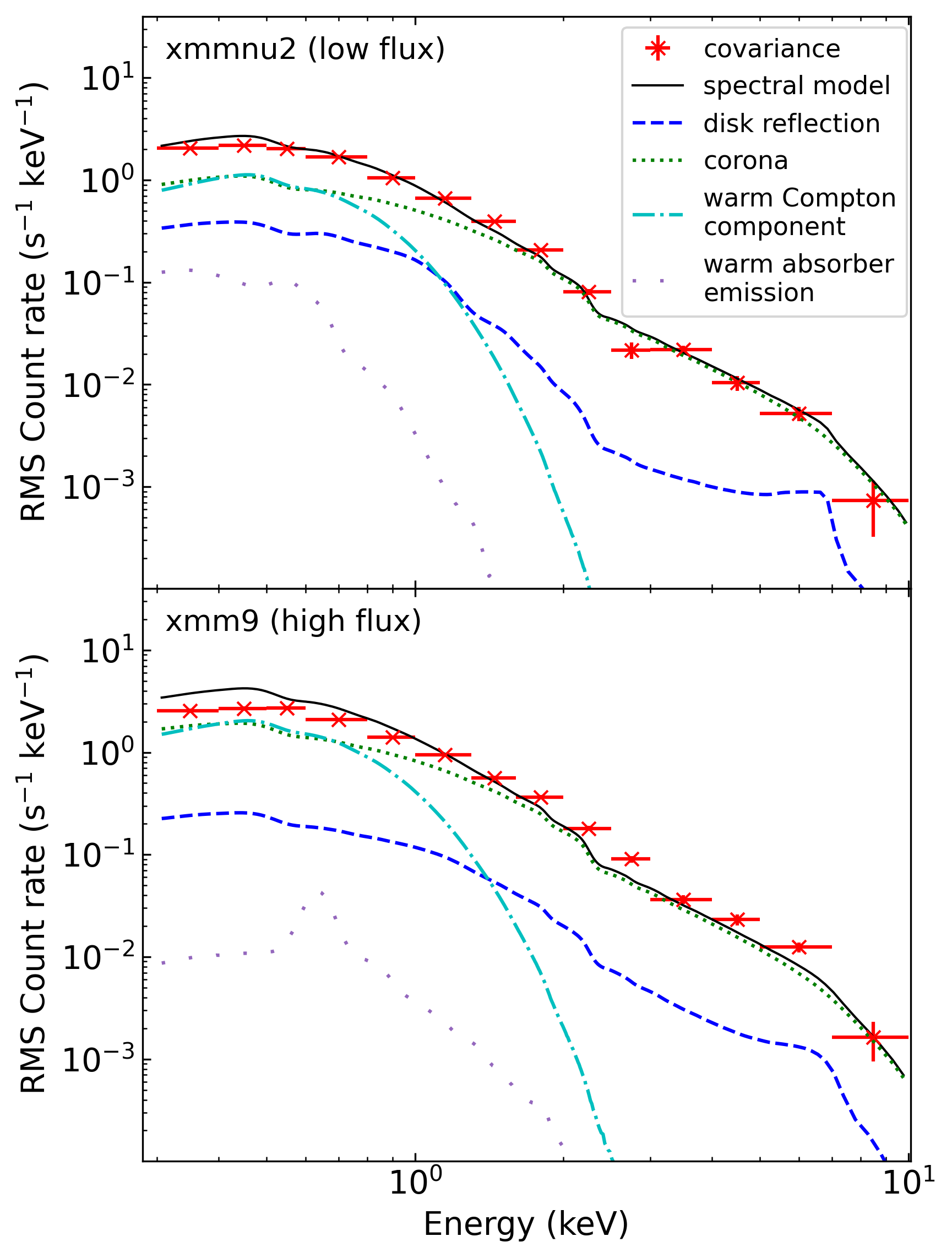}
\figcaption{Covariance spectra for a low-flux epoch xmmnu2 (upper panel) and a high-flux epoch xmm9 (lower panel) in comparison with the spectral model obtained from \textsc{XSPEC} fitting of the mean spectra. The red crosses represent the covariance spectra. The black solid line shows the best-fit spectral model of the mean spectra, while the blue dashed, green dotted, cyan dash-dotted, and purple loosely dotted lines correspond to the model components for the disk reflection, the coronal continuum, the warm Compton component, and the transmitted emission from the warm absorbers, respectively. The spectral models have been shifted along the y-axis so that the integrated count rate over the 0.3 - 10 keV range matches that of the covariance spectrum. \label{fig:cov}}
\end{figure}

Figure \ref{fig:cov} shows the covariance spectra for a low-flux epoch xmmnu2 and a high-flux epoch xmm9 alongside the spectral model obtained from \textsc{XSPEC} fitting of the mean spectra; similar plots for other epochs are presented in Figure \ref{fig:covall} in Appendix \ref{appsec:figs}. In our reverberation modeling, only the shape of the spectral model and the relative strengths of the components are important, while the absolute normalization is not critical. For ease of comparison, we shift the mean spectral model in each panel of Figures \ref{fig:cov} and \ref{fig:covall} along the y-axis so that the integrated count rate over the 0.3 - 10 keV range matches that of the covariance spectrum. For the low-flux epoch xmmnu2, the shape of the covariance spectrum is broadly consistent with the best-fit model of the mean spectra, with only minor discrepancies in the 0.45 keV and 2.75 keV energy bins. Most other epochs also exhibit general agreement between the covariance spectra and the mean spectral models. These results support the validity of using the mean spectral models in our reverberation modeling. 

In contrast, the covariance spectrum of the high-flux epoch xmm9 is noticeably flatter than the mean spectra, suggesting a deficiency of correlated variability in the soft bands. One potential explanation is that the warm Compton component, which dominates the soft band, contains a constant part. This constant part reduces its fractional variability and, consequently, its contribution to the covariance spectra, leading to the decrement in the covariance spectrum relative to the mean spectrum. As a result, the warm Compton component would have less contribution to the reverberation signal than assumed in Sections \ref{subsec:simcpt} - \ref{subsec:simthpl}. This will lead to a larger discrepancy between the model lag-energy spectra and observations, as the models become flatter due to a stronger dilution effect in the soft bands.

\subsection{Reverberation of a Slim Disk} \label{subsec:slimdisk}
\noindent In both the spectral fitting and reverberation modeling, our baseline model is a standard geometrically thin disk with a compact lamppost corona, plus a warm Comptonized layer on the disk surface to model the soft excess. Although this simplified scenario reasonably describes the flux spectra and most of the lag-energy spectra, it does not explain why the reverberation of the warm Compton component is only required in the lag-energy spectra of the high-flux epochs. In this section, we propose a reverberation model based on a slim disk to better explain the results presented in previous sections.

\begin{figure}[ht]
\epsscale{1.2}
\plotone{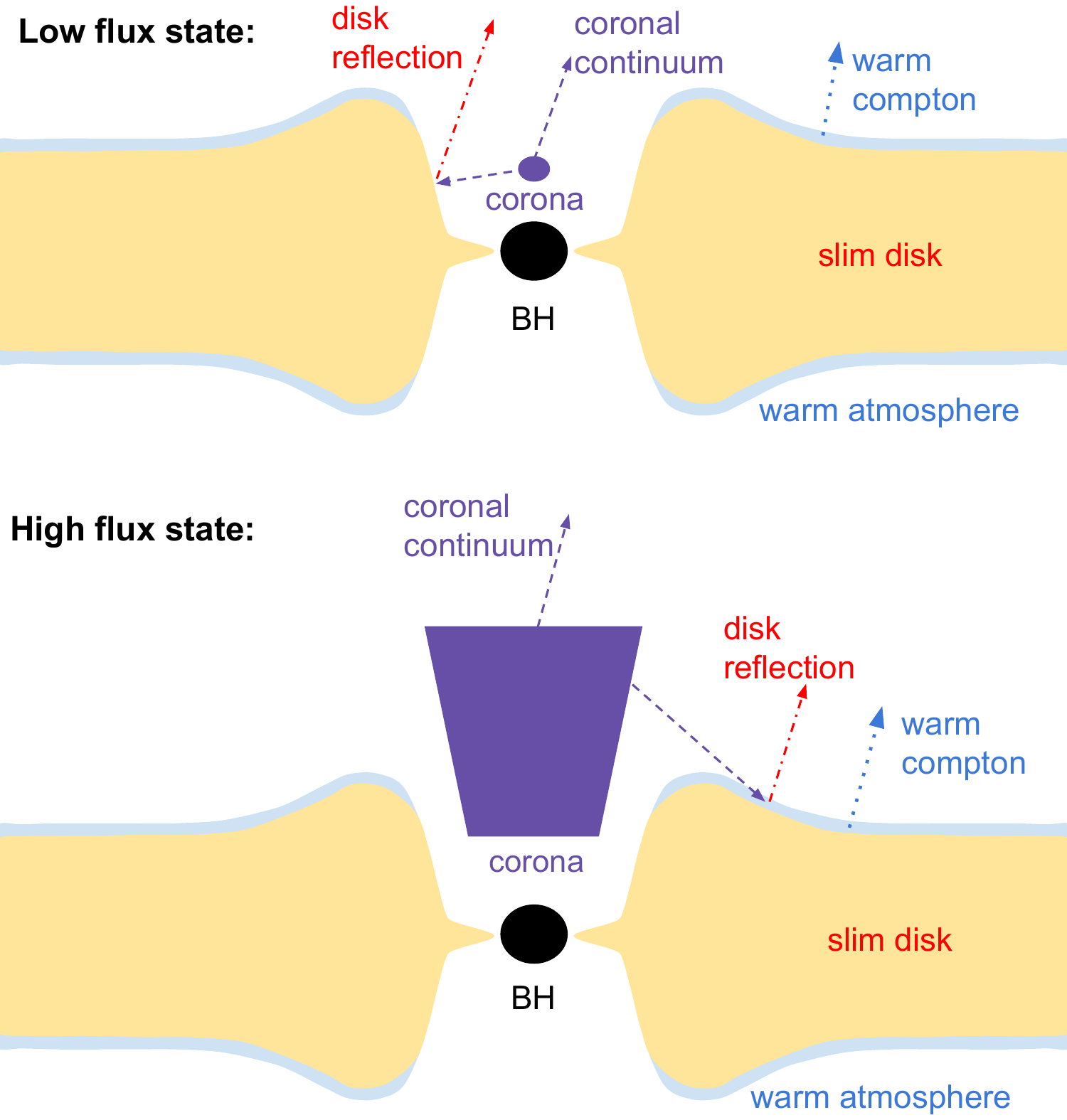}
\figcaption{Illustration of the slim disk scenario in the low-flux state (upper portion) and high-lux state (lower portion). In this scenario, the outer region of the accretion flow is puffed up to form a slim disk, and the warm atmosphere is a thin layer on the disk surface. In the low-flux state, the corona is located below the scale height of the accretion disk, preventing the coronal continuum from illuminating the warm atmosphere; consequently, the warm atmosphere does not contribute to the reverberation lag. In the high-flux state, however, the corona is elevated sufficiently to illuminate the outer disk and the warm atmosphere, causing the warm atmosphere to reverberate in response to the continuum variability and contribute to the lag signal. \label{fig:slimdisk}}
\end{figure}

Figure \ref{fig:slimdisk} presents schematic diagrams illustrating the slim disk scenario. In this model, the inner parts of the disk (in the regions beyond the innermost stable circular orbit) are puffed up \citep[e.g.,][]{McKinney2012}. The warm Comptonizing plasma remains as a thin layer on outer disk surface, akin to an atmosphere. In the low-flux state, the corona is located at a relatively low height, and the disk geometry prevents the corona illuminating the warm atmosphere. Consequently, the warm atmosphere does not interact directly with the coronal continuum and therefore does not contribute to the reverberation lag. This explains why the model lag-energy spectra agree with the observations in low-flux epochs without incorporating the warm atmosphere reverberation. Nevertheless, variability in the corona and the warm atmosphere may still be coupled through the propagation of fluctuations passing through the inner boundary of the accretion disk, allowing the warm atmosphere to contribute to the covariance spectra in the low-flux state. This accounts for the agreement between the covariance spectra and the mean spectra shown in the upper panel of Figure \ref{fig:cov}. 

In the high-flux state, the corona is located at a greater height, allowing it to illuminate the warm atmosphere. Under this condition, the warm atmosphere reverberates in response to continuum variability and contributes to the lag signal, consistent with the requirements for the warm atmosphere reverberation in the lag-energy spectra of high-flux epochs. In the outer regions of the disk, the variability of the warm atmosphere driven by the coronal continuum is smeared out over a broader range of time delay and therefore exhibits a lower amplitude than in the inner region. Because the warm atmosphere in the outer disk produces softer photons due to the lower temperature, its reduced variability leads to a deficiency at the soft end of the covariance spectra. This mechanism accounts for the discrepancy between the mean spectral models and the covariance spectra observed in high-flux epochs.  

In the slim disk scenario described above, we may further incorporate a geometrically extended corona instead of a compact lamppost. For a vertically extended corona, the apparent increase of corona height can also represent an upward expansion of the corona. In spectral fitting, the primary constraint on corona height comes from the emissivity profile of disk reflection. The emissivity profile of a slim disk illuminated by an extended corona likely differs from that of a thin disk illuminated by a compact lamppost as assumed in the spectral fitting. Consequently, the effective corona height derived from spectral fitting may not correlate well with the the light travel time between the corona and the disk. This is consistent with the lack of correlation between lag and corona height discussed in Section \ref{fig:lagtrend}.

A vertically extended corona may also reconcile discrepancies arsing from the potential deficiency of warm atmosphere variability in high-flux epochs. As discussed in Section \ref{subsec:cov}, lower variability leads to stronger dilution of the soft lags than assumed in Sections \ref{subsec:simcpt} - \ref{subsec:simthpl}, which flattens the model lag-energy spectra and increases the discrepancy with observations. To address this, we consider a corona with two vertically separated zones \citep[e.g.,][]{Chainakun2017}: a stable lower region that varies in concert with disk reflection and an upper outflow. When the extended corona is moving slowly at the base, strong light bending focuses rays from this part onto the inner disk. If the corona accelerates at larger heights, special relativistic beaming directs the continuum emitted from the top parts of the corona away from the disk. Consequently, additional variability generated at larger heights in the corona, possibly due to turbulence, will be decoupled between the continuum and disk reverberation. This decoupling reduces the effective dilution of the reverberation signal by the coronal continuum and increases the observed lag, since only a portion of the continuum contributes to reverberation. The variability of the upper corona still correlates with itself, so the covariance spectra are unaffected. 

The vertically extended corona could resemble an ``aborted jet'' that only propagates a short distance \citep[e.g.,][]{Ghisellini2004}. While some polarimetric studies based on the Imaging X-ray Polarimetry Explorer (IXPE) disfavors such a conical shape \citep[e.g.,][]{Tagliacozzo2023,Ingram2023,Gianolli2024}, those targets generally have relatively low Eddington ratios. In contrast, the coronae of high-Eddington-ratio AGN may have different geometries. For instance, spectral timing analysis of I Zwicky 1 suggests a combination of a radially extended corona and a vertically extended core that resembles the base of a jet \citep{Wilkins2017}.

The slim disk and extended corona scenario further explains the low reflection fraction derived from spectral fitting. In the slim disk geometry, coronal photons reflected by the inner disk can undergo additional reflection before escaping toward infinity. This multi-scattering reduces the strength of the disk reflection and leads to a low reflection fraction. In the high-flux state, the relativistic light bending effect becomes weaker, allowing more coronal photons to escape directly without being reflected. Moreover, the potential outflow of the corona at large heights discussed above can collimate the continuum emission away from the accretion disk \citep[e.g.,][]{Wilkins2015_reffrac}. These factors further reduce the reflection fraction at high fluxes and explain the anti-correlation between the reflection fraction and count rate observed in Figure \ref{fig:rffr}. 

The slim disk scenario also accounts for the high iron abundance and low corona temperature derived from spectral fitting. Multiple reflections of coronal photons by the inner disk enhance the iron line features in the reflection spectrum \citep[e.g.,][]{Ross2002,Wilkins2020}, leading to an increased apparent iron abundance. In the low-flux state, the corona is exposed to an enhanced number of UV photons from the inner disk relative to the thin disk scenario, leading to increased Compton cooling and therefore a lower corona temperature, as derived from spectral fitting of the low-flux epochs xmmnu1, xmmnu2 and nu1. The \textit{NuSTAR} observation analyzed by \citet{Kara2017} is also in a low-flux state, given its lower count rate compared to nu1, so the low corona temperature they measured is consistent with the slim disk scenario. 

Furthermore, the Fe K$\alpha$ line profile in a slim disk reflection spectrum differs from that of a thin disk. When the corona is located below the scale height of the disk, reflection is confined to the inner disk, enhancing the red wing of the Fe K$\alpha$ line due to stronger gravitational redshift in the vicinity of the SMBH than in the outer disk. In addition, the lack of reflection from outer disk weakens the blue wing of the Fe K$\alpha$ line. Moreover, the inner disk is expected to be highly ionized, which reduces the overall strength of the Fe K$\alpha$ line. Previous studies have developed reflection models of the slim disk \citep[e.g.,][]{Taylor2018,Shashank2024}, while their assumed geometries are relatively simple. Detailed spectroscopic and reverberation analyses based on these models are beyond the scope of this paper.

\section{Summary and Conclusion} \label{sec:summary}
\noindent We have performed a systematic spectral and timing analysis of Ark 564, one of the brightest X-ray AGN accreting near the Eddington limit. We used $\sim 500$ ks of \textit{XMM-Newton} observations and $\sim 400$ ks of \textit{NuSTAR} observations obtained over 13 years. We fit the X-ray spectra with a relativistic disk reflection model \textsc{relxilllpCp}, a coronal continuum model \textsc{nthcomp}, a warm Compton component \textsc{comptt} and three warm absorbers modeled with \textsc{XSTAR} tables. The inclusion of the warm Compton component \textsc{comptt} is essential for obtaining a good spectral fit.

We calculated the X-ray inter-band lags as a function of Fourier frequency and energy using a Fourier analysis approach. Assuming a lamppost geometry for the corona and a standard thin accretion disk, we combined ray-tracing simulations and the spectral model obtained from \textsc{XSPEC} fitting to generate models of the X-ray lag. Our main findings are as follows:
\begin{enumerate}
\item The X-ray lags vary over time. Soft lags in the high-flux epochs are substantially larger than those in the low-flux epochs, and the soft lag does not correlate with the corona height derived from spectral fitting. 
\item Fe K$\alpha$ lags are detected in only about 4 out of 11 epochs. The Fe K$\alpha$ lag does not correlate with X-ray flux, corona height, or the soft lag. 
\item The predicted lag-energy spectra by the disk reflection model broadly agree with most observations, except in the high-flux epochs. Reverberation of the warm Compton component accounts for the excess of soft lags in the high-flux epochs, except in the two softest energy bins of xmm9. 
\item A slim disk combined with an extend corona provides a better explanation for the evolution of the spectral parameters, time lags, and covariance spectra than the thin disk and lamppost corona model.  
\end{enumerate}

The variability in the soft lags and the intermittent detection of Fe K$\alpha$ lags raise concerns about drawing conclusions regarding the structure of the accretion disk and corona based solely on X-ray lags from a single epoch or from stacked observations. Multi-epoch spectral and timing analyses are therefore important for obtaining a comprehensive view of the accretion disk and corona in future studies. Resolving the time evolution of X-ray lags requires a large number of photons, making the effective area of the telescope critical for such studies. The large effective area of the future Advanced Telescope for High ENergy Astrophysics \citep[Athena, e.g.,][]{athena} will advance X-ray reverberation studies for probing the accretion disk and corona of AGN. \\

\noindent We thank the anonymous referee for helpful comments and suggestions. We thank Chloe Taylor for helpful discussions. We refined the English writing of this paper using the artificial intelligence tool ChatGPT (https://chatgpt.com/). This work was supported by the NASA Astrophysics Data Analysis Program (ADAP) under grant number 80NSSC22K0406. This work is based on observations obtained with \textit{XMM-Newton}, an ESA science mission with instruments and contributions directly funded by ESA Member States and NASA, and the \textit{NuSTAR} mission, a project led by the California Institute of Technology, managed by the Jet Propulsion Laboratory, and funded by NASA. Computing for this project was performed on the Sherlock cluster. The authors thank Stanford University and the Stanford Research Computing Center for providing computational resources and support.

%

\vspace{5mm}
\facilities{XMM-Newton(PN), NuSTAR}


\software{Astropy \citep{astropy_2013,astropy_2018}, SciPy \citep{scipy}, Numpy \citep{numpy}, Matplotlib \citep{matplotlib}, Pandas \citep{pandas}, XSPEC \citep{xspec}, pyLag (https://github.com/wilkinsdr/pyLag)}



\appendix

\section{Supplementary Figures} \label{appsec:figs}
\begin{figure*}
\gridline{\fig{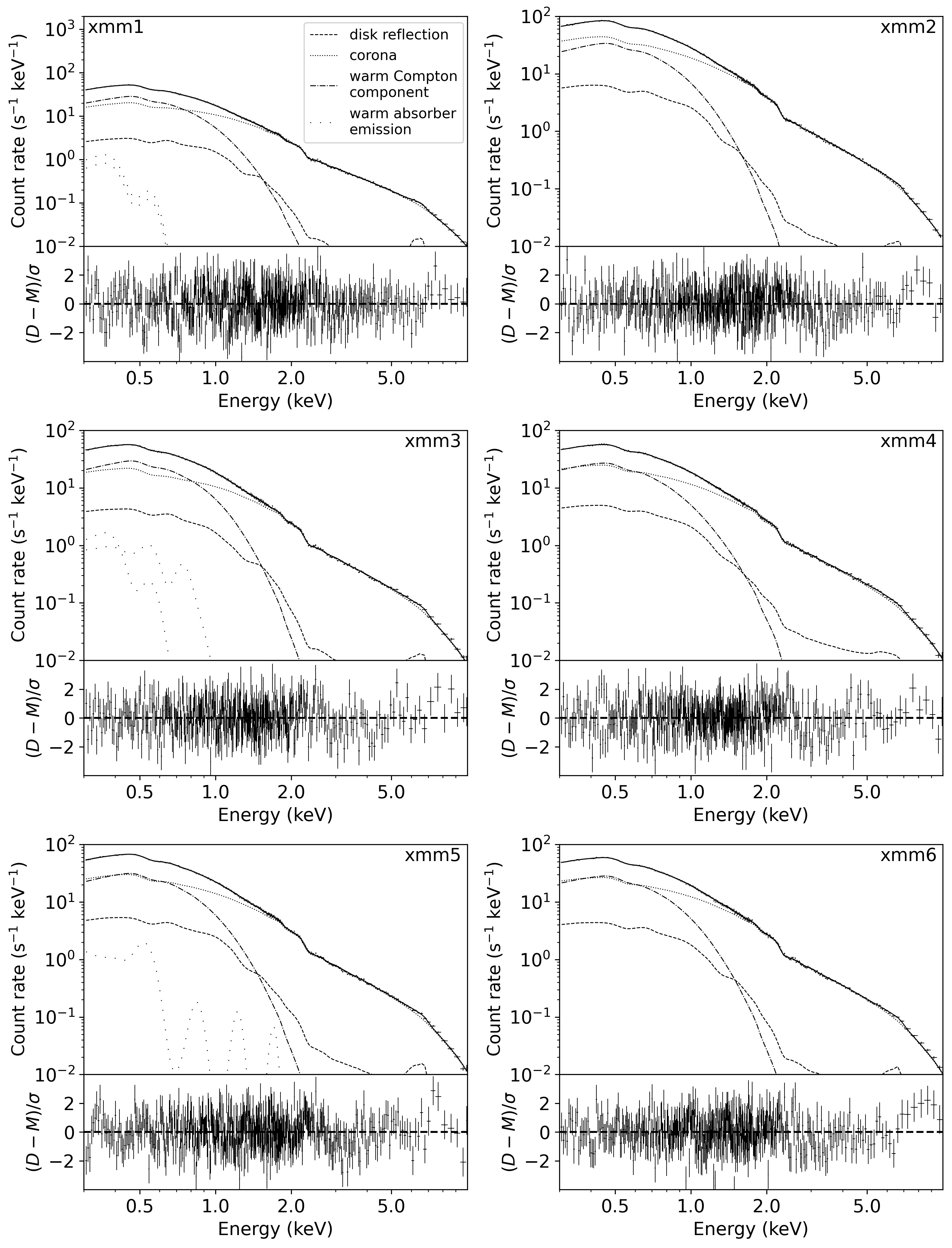}{0.93\textwidth}{}}
\figcaption{Same as Figure \ref{fig:specfit}, but for all other \textit{XMM-Newton} epochs that do not have simultaneous \textit{NuSTAR} observations. The epoch name is given in the upper left or right corner of each panel. \label{fig:specfit_all0}}
\end{figure*}

\begin{figure*}
\gridline{\fig{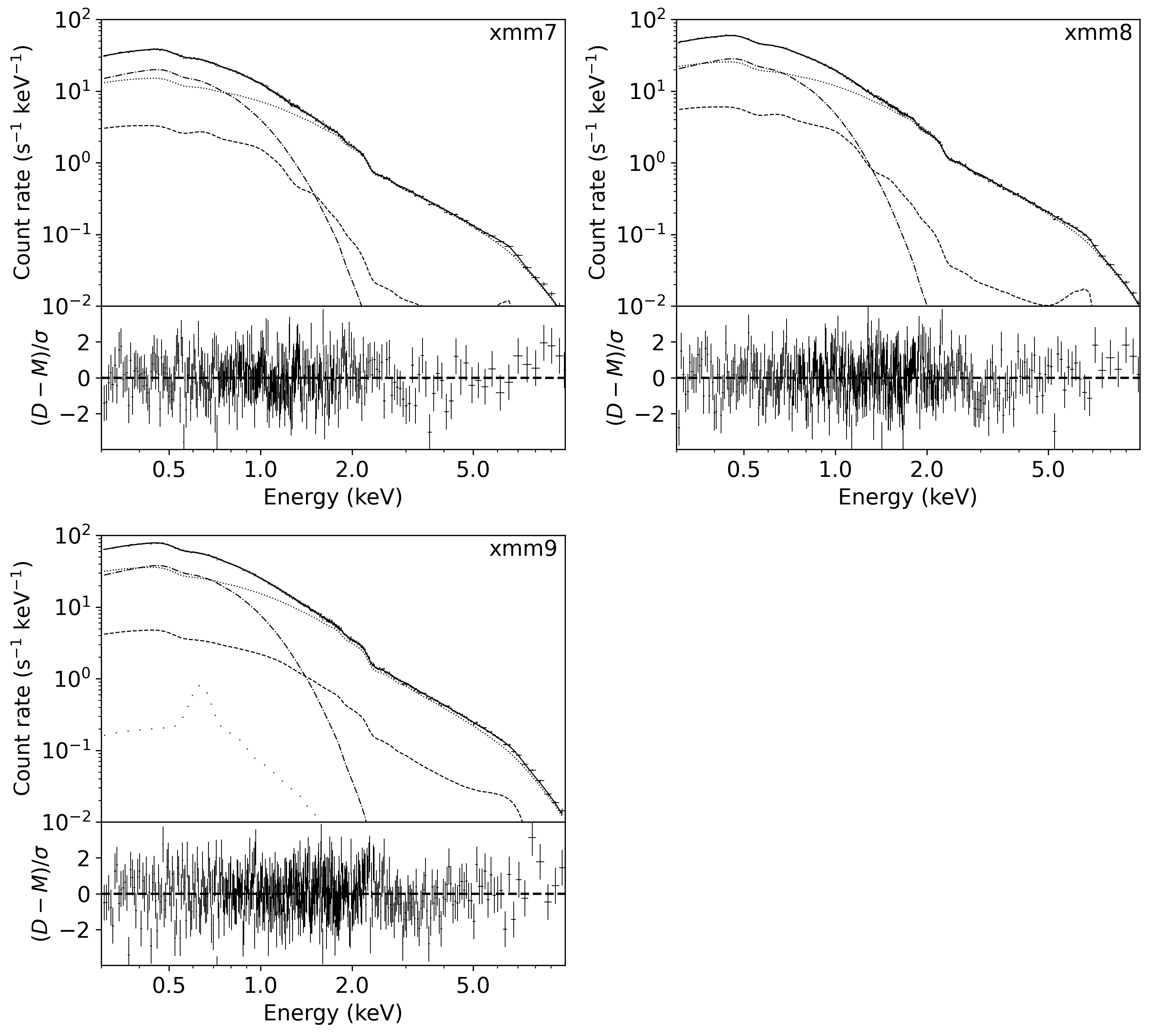}{0.93\textwidth}{}}
\figcaption{Figure \ref{fig:specfit_all0}, continued \label{fig:specfit_all1}}
\end{figure*}

\begin{figure*}
\gridline{\fig{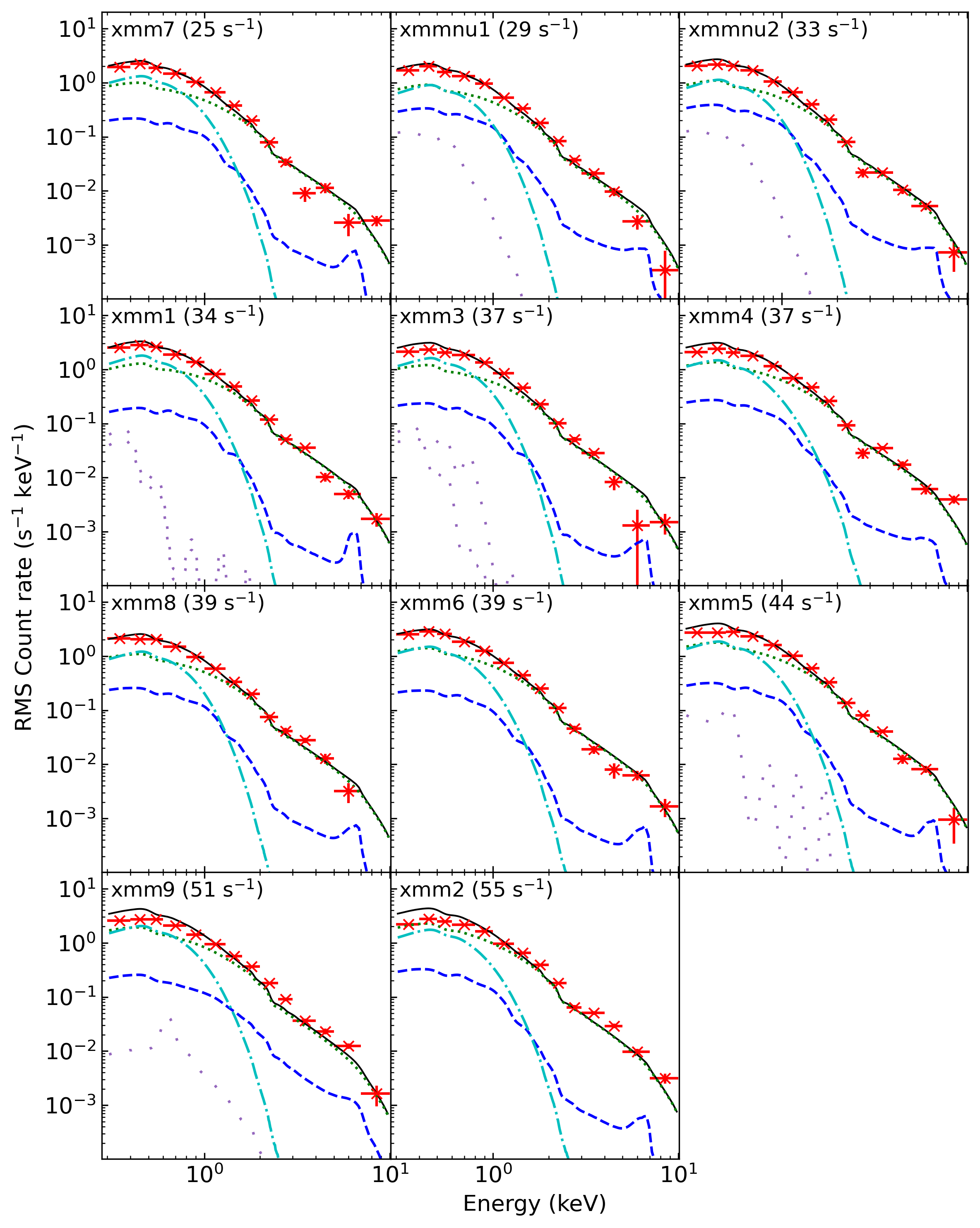}{0.93\textwidth}{}}
\figcaption{Same as Figure \ref{fig:cov}, but for all \textit{XMM-Newton} epochs. Each panel is for an epoch with its name and count rate given in the upper left corner. \label{fig:covall}}
\end{figure*}

We present the supplementary figures in this section. Figures \ref{fig:specfit_all0} and \ref{fig:specfit_all1} show the spectral fitting plots discussed in Section \ref{sec:spec} for all all \textit{XMM-Newton} epochs that are not shown in Figure \ref{fig:specfit}. Figure \ref{fig:covall} shows the covariance spectra discussed in Section \ref{subsec:cov} for all \textit{XMM-Newton} epochs.

\bibliography{ref}{}
\bibliographystyle{aasjournal}



\end{document}